\documentclass[10pt,twocolumn,a4paper]{IEEEtran}

\usepackage[active]{srcltx} 
\usepackage{booktabs}
\usepackage[cmex10]{amsmath}
\usepackage{amsfonts}
\usepackage[final]{graphics}
\usepackage[final]{graphicx}
\usepackage{amsbsy}
\usepackage{amssymb}
\usepackage{url}
\usepackage{enumerate}
\usepackage{cite}
\usepackage{psfrag}
\usepackage{color}
\usepackage{euscript}
\usepackage[utf8]{inputenc}
\usepackage{algorithmic}
\usepackage[caption=false,font=footnotesize]{subfig}
\usepackage{tabularx}
\usepackage{float}
\usepackage[ruled,vlined]{algorithm2e}
\usepackage{stfloats}

\newcommand{\Cset}{\mathbb{C}}

\newcommand{\diag}{{\text{diag}}}

\newcommand{\eqdef}{\triangleq}

\newcommand{\herm}{\text{H}}
\newcommand{\trasp}{\text{T}}

\renewcommand{\vec}{\text{vec}}

\def\bdm#1\edm{\begin{displaymath}#1\end{displaymath}}
\def\be#1\ee{\begin{equation}#1\end{equation}}
\def\barr#1\earr{\begin{align}#1\end{align}}

\begin{document}

\title{
Gridless Full-Space DOA Estimation for STAR-RIS-Assisted Wireless Systems
}
\author{Ziming~Liu,~\IEEEmembership{Member,~IEEE},
Tao~Chen,~\IEEEmembership{Member,~IEEE},
Muran~Guo,~\IEEEmembership{Member,~IEEE} \\
and Francesco~Verde,~\IEEEmembership{Senior Member,~IEEE}
\thanks{
Manuscript received June 8, 2026; 
revised xx yy, 2026;
accepted xx yy, 2026.
The associate editor coordinating the review of this paper and
approving  it for publication  was Dr.~xx yy.
(\em Corresponding author: Francesco~Verde)
}
\thanks{
T.~Chen, Z.~Liu and M.~Guo are with Harbin Engineering University, Harbin,
150001 China (e-mail:chentao@hrbeu.edu.cn, lzmfred@hrbeu.edu.cn and guomuran@hrbeu.edu.cn).
F.~Verde is with the Department of Engineering, 
University of Campania Luigi Vanvitelli, Aversa I-81031, Italy
(e-mail: francesco.verde@unicampania.it).
}
}
%


\maketitle

\begin{abstract}
Simultaneously transmitting and reflecting reconfigurable intelligent 
surfaces (STAR-RIS) enable full-space ($0^\circ$--$360^\circ$) signal 
coverage, making them a compelling platform for integrated sensing and 
communication in next-generation wireless networks. In this paper, we 
investigate gridless direction-of-arrival (DOA) estimation across the 
full spatial domain in STAR-RIS-assisted systems operating with a single 
RF sensing chain. We show that the coupled reflection-transmission mechanism of STAR-RIS induces a multichannel finite-rate-of-innovation 
(FRI) structure in the received signal, which enables casting DOA 
estimation as a structured low-rank recovery problem without angular grid 
discretization. Building on this observation, we develop a proximal 
gradient descent algorithm with alternating projections onto a 
block-Hankel matrix set, enabling robust angle retrieval from limited 
measurements. Two practically relevant STAR-RIS configurations are 
addressed: element-wise uniform and nonuniform energy-splitting designs, 
each handled through a dedicated lifting strategy that preserves the 
underlying algebraic structure. A Ziv-Zakai bound is derived for the 
coupled full-space sensing model as a performance benchmark across the 
full SNR range. Numerical results show that the proposed methods 
consistently outperform grid-based baselines, achieving sub-degree 
accuracy within $\pm 60^\circ$ of boresight at comparable or lower 
computational cost.
\end{abstract}

\begin{IEEEkeywords}
Finite rate of innovation (FRI),
full-space angle estimation,
reconfigurable intelligent surface (RIS), 
simultaneously transmitting and reflecting (STAR) metasurface. 
\end{IEEEkeywords}

\section{Introduction}

Accurate angle estimation is a key enabler for beam management, user 
localization, and environment awareness in next-generation wireless 
networks. Within integrated sensing and communication (ISAC) systems, 
reliable direction-of-arrival (DOA) estimation plays a central role, 
particularly in challenging propagation environments such as dense urban 
scenarios, where blockage and rich scattering lead to severe 
non-line-of-sight (NLOS) conditions~\cite{S2024IRS,Z2022Locaion,xu2023spatial,yu2024dual,xu2023channel}. 
These impairments significantly degrade the performance of conventional sensing 
and localization techniques, motivating the need for new architectures 
that can enhance spatial observability.

Reconfigurable intelligent surfaces (RISs) have recently emerged as a 
promising technology to improve wireless sensing and communication by 
enabling programmable control of electromagnetic wave propagation. By 
dynamically adjusting the response of their constituent elements, RISs 
can reshape the propagation environment, enhance signal quality, and 
improve sensing robustness~\cite{Boccia2012Multilayer,J2012Wideband,liu2023snr,wang2025enhanced,pepe2026conformal}. 
However, conventional 
reflection-only RIS architectures are inherently limited to a single 
hemispherical region ($0^\circ$--$180^\circ$), which restricts their 
ability to provide full spatial coverage. Extending such architectures to 
full-space operation typically requires multiple coordinated surfaces and 
careful calibration~\cite{megahed2025deep,Ma2022Cooperative}, resulting in increased hardware 
complexity and deployment cost.

Simultaneously transmitting and reflecting RIS (STAR-RIS) overcomes this 
limitation by enabling concurrent transmission and reflection through 
multi-layer metasurface designs~\cite{ahmed2023survey,mu2021simultaneously}. 
STAR-RISs naturally support full-space ($0^\circ$--$360^\circ$) coverage and can 
simultaneously serve users located on both sides of the 
surface~\cite{xu2021star,xue2024noma,Verde.2024.SPM}, making them a powerful platform for 
full-space sensing and localization. Recent works have explored STAR-RIS 
for ISAC applications: multi-user mmWave systems have been investigated 
using time-slot switching or mode-coding strategies to enable cascaded 
channel estimation under limited pilot resources~\cite{luo2025Channel,yue2023Simultaneously}; 
Cram\'{e}r-Rao bounds have been derived for three-dimensional 
localization~\cite{He2023STAR,He2022Simultaneous}; and sensing-derived information such as 
position and velocity has been leveraged to enhance communication 
performance~\cite{Li2024STAR,Meng2024Sensing}. These results, together with ongoing 
standardization efforts~\cite{ETSI_GR_RIS_006}, demonstrate the growing relevance of 
STAR-RIS-enabled sensing.

\IEEEpubidadjcol

Despite these advances, dedicated DOA estimation methods tailored to 
STAR-RIS remain scarce, and the structural properties specific to 
STAR-RIS systems have not been fully exploited. The simultaneous presence 
of reflected and transmitted signal components gives rise to a coupled 
full-space observation model that differs fundamentally from classical 
single-hemisphere array processing. Conventional DOA estimators operating 
on one subspace at a time treat the contribution from the opposite 
subspace as unmodeled interference, introducing a structured bias that 
neither increasing the SNR nor refining the angular search grid can 
remove. An improved approach must therefore explicitly account for this 
reflection-transmission coupling from the outset.

\subsection{Prior work}

Classical subspace-based methods such as MUSIC and 
ESPRIT~\cite{sch1986MUSIC,Roy1986ESPRIT} rely on multiple snapshots collected under a fixed 
array manifold. In STAR-RIS-assisted sensing, however, each time slot 
typically corresponds to a different metasurface configuration, resulting 
in time-varying sensing matrices rather than repeated observations of the 
same array response. This lack of stationarity limits the applicability 
of covariance-based approaches and may require a large number of 
measurements to achieve reliable estimation.

Sparse reconstruction methods provide an alternative by exploiting 
angular-domain sparsity through compressive sensing, including on-grid 
approaches~\cite{zuo2021doa,TroppGilbert2007} and off-grid refinements such as sparse 
Bayesian learning (SBL)~\cite{Gerstoft2016SBL,jin2023off}. While these methods accommodate 
time-varying sensing matrices and have been applied to RIS-assisted 
systems~\cite{Lin2021Single,wang2022jointbeamformingdesign3d}, they suffer from a bias-complexity tradeoff: 
grid discretization introduces mismatch errors, and, when applied to 
STAR-RIS systems without accounting for the coupled full-space model, 
they incur a persistent high-SNR error floor driven by unmodeled 
inter-subspace interference.

Gridless approaches such as atomic norm minimization (ANM) avoid 
discretization errors by exploiting structured signal models. However, 
most existing formulations rely on Toeplitz or Hermitian structures 
induced by conventional array geometries. In STAR-RIS systems, the 
coupling between transmission and reflection responses, together with 
spatially varying energy-splitting coefficients, distorts these 
structures, making existing gridless methods less directly applicable. 
Recent efforts have considered simplified STAR-RIS configurations with 
element-wise uniform parameterizations~\cite{li2024simultaneously} or have proposed 
ADMM-based solvers under similar assumptions~\cite{chen2022efficient,li2023joint,chen2021reconfigurable,Chen2026Efficient}. 
However, a unified gridless framework that rigorously accounts for both uniform and 
nonuniform energy-splitting designs while explicitly exploiting the 
coupled full-space structure remains an open problem.

\subsection{Proposed approach and contributions}

In this paper, we address full-space DOA estimation in STAR-RIS-assisted 
wireless systems with a single RF sensing chain --- a hardware constraint 
motivated by the need to minimize receiver cost and complexity in 
large-scale deployments. The central insight is that the coupled 
transmission-reflection mechanism of STAR-RIS induces a 
finite-rate-of-innovation (FRI) structure~\cite{pan2016towards,simeoni2020cpgd,pan2018efficient,wang2021two}
in the  received signal: the aggregated metasurface output is a superposition of 
complex exponentials indexed by the unknown DOAs, which naturally admits 
a low-rank block-Hankel representation. This algebraic structure makes 
the problem amenable to gridless recovery via structured matrix methods, 
without requiring angular discretization or covariance estimation across 
stationary snapshots.

Building on this observation, we develop a proximal gradient descent 
(PGD) framework combined with alternating projections onto a block-Hankel 
matrix set to recover the latent FRI vector from compressed measurements, 
after which DOAs are extracted via annihilating-filter root finding. Our 
main contributions are as follows:

\begin{enumerate}

\item \textit{STAR-RIS-enabled full-space sensing model:} We exploit the 
intrinsic coupling between transmission and reflection responses to 
establish a unified full-space sensing model. Both element-wise uniform 
and nonuniform energy-splitting (ES) configurations are considered, and 
the corresponding structured multichannel FRI representations are 
derived.

\item \textit{Gridless DOA estimation via structured FRI modeling:} We 
show that the sensing model admits an FRI structure enabling 
gridless angle estimation from limited measurements. A structured 
low-rank recovery approach based on alternating projections and proximal 
gradient methods is developed, and sufficient local stability conditions
on the step-size 
and lifting parameter for convergence are established.

\item \textit{Extension to nonuniform STAR-RIS configurations:} For 
nonuniform ES settings, we introduce a block-structured lifting strategy 
that preserves the underlying algebraic structure and enables reliable 
angle recovery under more general conditions. We show that this 
formulation also resolves the high-SNR bias floor that affects methods 
neglecting the coupled full-space structure.

\item \textit{Performance bounds:} We derive a Ziv--Zakai bound (ZZB) 
for the coupled full-space sensing model, which accounts for 
the cross-subspace coupling induced by simultaneous transmission and 
reflection, and use it to assess the performance limits of the proposed 
methods.

\end{enumerate}

\subsection{Notation}

Boldface lowercase and 
uppercase letters denote vectors and matrices, respectively. 
Superscripts $(\cdot)^\trasp$, $(\cdot)^*$, and $(\cdot)^\herm$ denote 
transpose, complex conjugate, and Hermitian transpose. $\mathbf{I}_n$ 
is the $n\times n$ identity matrix, $\mathbf{0}_n$ the zero vector, 
and $\mathbf{1}_n$ the all-ones vector. $\|\cdot\|_2$ and 
$\|\cdot\|_F$ are the vector $\ell_2$ and matrix Frobenius norms. 
$\text{diag}[\mathbf{a}]$ is the diagonal matrix formed from vector 
$\mathbf{a}$; $\text{diag}[\mathbf{A}]$ extracts the diagonal of 
matrix $\mathbf{A}$. The symbols $\otimes$ and $\odot$ denote the 
Kronecker and Hadamard products, and $\oslash$ element-wise division 
between conformable diagonal matrices. $\text{vec}(\mathbf{A})$ stacks 
the columns of $\mathbf{A}$ into a vector. $\text{rank}[\mathbf{A}]$ 
and $\ker(\mathbf{A})$ denote the rank and null space of $\mathbf{A}$, 
and $\sigma_i(\mathbf{A})$ its $i$-th singular value.

\subsection{Paper organization}

The remainder of this paper is organized as follows. Section~\ref{sec:system} presents the system model. 
Section~\ref{sec:Uniform} develops the proposed method for uniform STAR-RIS configurations, while Section~\ref{sec:Nonuniform} extends the framework to 
nonuniform settings. Section~\ref{sec:Perform} discusses performance aspects, and Section~\ref{sec:results} reports numerical results. 
Section~\ref{sec:conclusions} concludes the paper.

\section{System model and basic assumptions}
\label{sec:system}

We consider an urban uplink scenario where users are located both 
outdoors and indoors. A sensing module equipped with a single radio 
frequency (RF) chain at the base station (BS) collects uplink signals 
and estimates the angles of all users.
In dense urban deployments, the line-of-sight path from outdoor users to 
the BS may be blocked by surrounding buildings, while indoor users 
suffer additional penetration loss. To compensate for blockage and 
simultaneously serve both user groups, a STAR-RIS is mounted on the 
building facade. Using the STAR-RIS plane as a reference boundary, the 
full-space domain is partitioned into two half-spaces: the reflection 
space (RS), on the same side as the BS (outdoor users), and the 
transmission space (TS), on the opposite side (indoor users). 
Specifically, $K_R$ users are located in the RS and $K_T$ users in the 
TS, for a total of $K \triangleq K_R + K_T$ users.

We adopt the narrowband signal assumption, where the uplink bandwidth 
$\Delta f$ satisfies $\Delta f \ll f_0$, with $f_0$ being the carrier 
frequency. Without loss of generality, the STAR-RIS is a uniform linear 
array (ULA) with $N > K$ elements spaced $\lambda/2$ apart, where 
$\lambda = c/f_0$ and $c$ is the speed of light.

\begin{figure}[t]
	{\includegraphics[width=\linewidth]{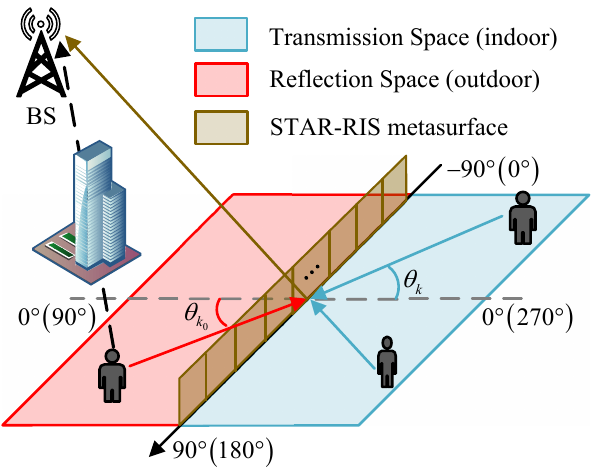}}
	\caption{Schematic diagram of the STAR-RIS-assisted system.}
	\label{fig:space_angle}
\end{figure}

Throughout the paper, two angle conventions are used depending on 
context. In the \emph{full-space} representation, the metasurface plane 
is aligned with the $0^\circ$--$180^\circ$ axis, with the surface normal 
pointing into the RS defined as $90^\circ$ and the normal pointing into 
the TS as $270^\circ$. In the \emph{semi-space} representation, each 
subspace is independently parameterized from $-90^\circ$ to $90^\circ$, 
where $0^\circ$ is perpendicular to the metasurface in the corresponding 
subspace. The considered geometry is illustrated in Fig.~\ref{fig:space_angle}, 
where angles outside brackets follow the semi-space convention and angles 
inside brackets the full-space convention.

\subsection{STAR-RIS model}

The STAR-RIS coefficients are updated on a slot-by-slot basis, yielding 
a time-varying effective sensing matrix that increases measurement 
diversity and improves identifiability of the angle parameters, 
particularly with a single RF chain. We assume the reconfiguration is 
sufficiently fast that user angles are quasi-static over $T_\text{s}$ slots, 
while the control sequence is known at the BS.
The STAR-RIS operates in energy-splitting (ES) mode, where the 
electromagnetic response of the $n$-th element during slot $t$ is 
described by the transmission and reflection coefficients
\begin{equation}
    \phi_{i,n}(t) = \beta_{i,n}(t)\, e^{j\varphi_{i,n}(t)}, 
    \quad \text{ for $i \in \{\text{R}, \text{T}\}$}
    \label{eq:coeff}
\end{equation}
where $0 < \beta_{i,n}(t) \leq 1$ and $\varphi_{i,n}(t) \in [0, 2\pi)$ 
are the amplitude and phase of the $n$-th element, for 
$n \in \mathcal{N} \triangleq \{1,\ldots,N\}$ and 
$t \in \mathcal{T} \triangleq \{1,\ldots,T_\text{s}\}$.

In ES mode, the passive lossless STAR-RIS must satisfy two constraints. 
Power conservation requires
\begin{equation}
    \beta_{\text{R},n}^2(t) + \beta_{\text{T},n}^2(t) = 1
    \label{eq:power}
\end{equation}
and the purely reactive impedance of each element imposes the 
phase-difference constraint
\begin{equation}
    \Delta\varphi_n(t) \triangleq \varphi_{\text{R},n}(t) - \varphi_{\text{T},n}(t) 
    = \tfrac{\pi}{2} \;\text{ or }\; \tfrac{3\pi}{2}.
    \label{eq:phase}
\end{equation}
The STAR-RIS is collectively described by the diagonal matrices 
$\boldsymbol{\Phi}_i(t) \in \mathbb{C}^{N\times N}$, with $i \in \{\text{T},\text{R}\}$, 
defined as
\begin{equation}
    \boldsymbol{\Phi}_i(t) \triangleq 
    \text{diag}\!\left[\beta_{i,1}(t)e^{j\varphi_{i,1}(t)},\ldots,
    \beta_{i,N}(t)e^{j\varphi_{i,N}(t)}\right].
    \label{eq:Phi}
\end{equation}
Constraint~\eqref{eq:phase} implies $e^{j\varphi_{\text{R},n}(t)} = 
\pm j\, e^{j\varphi_{\text{T},n}(t)}$, so that the reflection and 
transmission coefficients of the $n$-th element are related by
\begin{equation}
    \beta_{\text{R},n}(t)\,e^{j\varphi_{\text{R},n}(t)} = 
    \pm j\sqrt{1 - \beta_{\text{T},n}^2(t)}\; e^{j\varphi_{\text{T},n}(t)}
    \label{eq:coupling}
\end{equation}
where the sign is element- and slot-dependent. Equation~\eqref{eq:coupling} 
establishes a deterministic mapping between $\boldsymbol{\Phi}_\text{R}(t)$ and 
$\boldsymbol{\Phi}_\text{T}(t)$: given one matrix, the other is fully determined 
up to the binary sign choice. In matrix form,
\begin{equation}
    \boldsymbol{\Phi}_\text{R}(t) = 
    \left[\mathbf{I}_N - \mathbf{B}_\text{T}^2(t)\right]^{1/2} 
    \oslash \mathbf{B}_\text{T}(t) \odot \mathbf{J}(t) \odot \boldsymbol{\Phi}_\text{T}(t)
    \label{eq:mapping}
\end{equation}
where $\mathbf{J}(t) \triangleq \text{diag}[j_1(t),\ldots,j_N(t)]$ 
with $j_n(t) = \pm j$, and 
$\mathbf{B}_\text{T}(t) \triangleq \text{diag}[\beta_{\text{T},1}(t),\ldots,\beta_{\text{T},N}(t)]$.

\subsection{Signal model at the base station}

All users are assumed to be in the far field of the STAR-RIS, with 
azimuth angles $\theta_1,\ldots,\theta_K$ relative to the metasurface 
array. The steering vector of the $k$-th user is
\begin{equation}
    \mathbf{a}(\theta_k) \triangleq 
    \left[1,\, e^{-j\pi\sin\theta_k},\, \ldots,\, 
    e^{-j\pi(N-1)\sin\theta_k}\right]^\trasp.
    \label{eq:steering}
\end{equation}

During the sensing stage, pilot-aided signaling ensures that the 
complex gain $s_k(t)$ of the $k$-th user can be treated as 
slot-invariant, i.e., $s_k(t) \equiv s_k$~\cite{chen2022efficient,zheng2024ris}. 
Let 
$\mathcal{K}_\text{R} \eqdef \{1,2,\ldots,K_\text{R}\}$ and
$\mathcal{K}_\text{T} \eqdef \{K_\text{R}+1,K_\text{R}+2,\ldots,K\}$,  
the impinging signal at the STAR-RIS in each slot reads as
\be
\mathbf{x}_i= \sum\limits_{k \in \mathcal{K}_i} \mathbf{a}(\theta_k) \,  {s_k}  
= {\bf{A}}_i \, \mathbf{s}_i \:, 
\,\, \text{for $i\in \left\{ \text{T},\text{R} \right\}$}
\label{eq:x}
\ee
for $t \in \mathcal{T}$, with  
\barr
\mathbf{A}_\text{R} & \eqdef \left[ \mathbf{a}(\theta_1), \ldots, \mathbf{a}(\theta_{K_\text{R}})\right] \in \Cset^{N \times K_\text{R}} 
\\
\mathbf{A}_\text{T} & \eqdef \left[ \mathbf{a}(\theta_{K_\text{R}+1}), \ldots, \mathbf{a}(\theta_K)\right] \in \Cset^{N \times K_\text{T}} 
\\
\mathbf{s}_\text{R} & \eqdef [s_1, \ldots, s_{K_\text{R}}]^\trasp \in \Cset^{K_\text{R}}
\\
\mathbf{s}_\text{T} & \eqdef [s_{K_\text{R}+1}, \ldots, s_K]^\trasp \in \Cset^{K_\text{T}} \: .
\earr

The signal collected by the BS at slot $t$ is
\begin{equation}
y(t)= \mathbf{h}^\trasp \left[{{{\bf{\Phi }}_{\text{R}}}(t) \, {{\bf{x}}_{\text{R}}}
+ {{\bf{\Phi }}_{\text{T}}}(t) \, {{\bf{x}}_{\text{T}}}} \right] + n(t)
\label{data_BS2}
\end{equation}
where $\mathbf{h} \in \mathbb{C}^N$ is the time-invariant channel 
between the STAR-RIS and the BS, assumed known, and 
$n(t) \sim \mathcal{CN}(0,\sigma_n^2)$ is additive white Gaussian noise.

Exploiting the deterministic mapping~\eqref{eq:mapping}, the received signal 
can be written using $\boldsymbol{\Phi}_\text{R}(t)$ alone as
\barr
y(t) & = \mathbf{h}^\trasp  {{{{\bf{ \Phi }}}_{\text{R}}(t)} 
\sum\limits_{k = 1}^K {\bf{\tilde a}}\left(t; {{\theta _k}} \right) {s_k}}  + n(t)
\nonumber \\ & = \mathbf{h}^\trasp  {{{\bf{ \Phi }}}_{\text{R}}(t)} \, \tilde{\mathbf{A}}(t) 
\, \mathbf{s} + n(t)
\label{eq:y-final}
\earr
with ${\bf{\tilde a}}\left(t; {{\theta _k}} \right) \eqdef {{\bf{j}}_k}(t) \odot {{\bf{p}}_k}(t) 
\odot {\bf{a}}\left( {{\theta _k}} \right) \in \Cset^N$, 
where ${{\bf{j}}_k}(t)$ and ${{\bf{p}}_k}(t)$ are the phase and amplitude correction vectors, 
respectively, which
are defined as 
\be
{{\bf{j}}_k}(t)  \eqdef \left\{ {\begin{array}{*{20}{c}}
{{\bf{1}}_N}\:, &{\text{for $k \in \mathcal{K}_\text{R}$}}\:; \\
{{{\left[ {{j_1}(t),\ldots ,{j_N}}(t) \right]}^\trasp}}\:, &{\text{for $k \in \mathcal{K}_\text{T}$}}\:;
\end{array}} \right.
\label{eq:j}
\ee
\be{{\bf{p}}_k}(t) \eqdef \left\{ {\begin{array}{*{20}{c}}
{{\bf{1}}_N}\:, &{\text{for $k \in \mathcal{K}_\text{R}$}}\:;\\
{{{\left[ {{p_1}(t),\ldots ,{p_N}(t)} \right]}^\trasp}}\:, &{\text{for $k \in \mathcal{K}_\text{T}$}}\:;
\end{array}} \right.
\label{eq:p}
\ee
with $p_n(t) \eqdef [1-\beta^2_{\text{R},n}(t)]^{1/2}/\beta_{\text{R},n}(t)$, and, finally, we
have defined the matrix 
$\tilde{\mathbf{A}}(t) \eqdef [{\bf{\tilde a}}\left(t; {{\theta_1}} \right), \ldots, {\bf{\tilde a}}\left(t; {{\theta_K}} \right)] \in \Cset^{N \times K}$,
for $t \in \mathcal{T}$, and the vector $\mathbf{s} \eqdef [\mathbf{s}_\text{R}^\trasp,\mathbf{s}_\text{T}^\trasp]^\trasp \in \Cset^K$.

Model~\eqref{eq:y-final} highlights that the received signal is a 
superposition of reflection- and transmission-space contributions under 
a common STAR-RIS-aided observation. When one subspace is processed in 
isolation, the contribution from the opposite subspace appears as an 
unmodeled structured term rather than independent noise, potentially 
introducing model mismatch and reducing the effective SNR. Accordingly, 
properly exploiting the intrinsic reflection--transmission coupling is 
essential for reliable full-space angle estimation.

\section{Full-Space DOA Estimation Under Element-Wise Uniform STAR-RIS 
Parameters}
\label{sec:Uniform}

We first consider the \emph{element-wise uniform} STAR-RIS regime, where 
all metasurface elements share identical parameters, i.e., 
$\beta_{i,n}(t)=\beta_i(t)$ and $\Delta\varphi_n(t)=\Delta\varphi(t)$, 
for all $n \in \mathcal{N}$ and $t \in \mathcal{T}$. In this case, 
by virtue of \eqref{eq:j} and \eqref{eq:p}, the matrix 
$\widetilde{\mathbf{A}}(t)$ partitions as
\begin{equation}
\widetilde{\mathbf{A}}(t) = \left[\mathbf{A}_\text{R},\; g(t) \mathbf{A}_\text{T}\right]
\label{eq:Atildepart-unif}
\end{equation}
with $g(t) \triangleq \pm j\,[1-\beta^2_\text{R}(t)]^{1/2}/\beta_\text{R}(t)$. 
Substituting \eqref{eq:Atildepart-unif} into \eqref{eq:y-final} gives the 
equivalent signal model
\begin{equation}
y(t) = \mathbf{h}^\trasp \boldsymbol{\Phi}_\text{R}(t)\,\mathbf{r}(t) + n(t)
\label{eq:y-final-2}
\end{equation}
where $\mathbf{r}(t) \triangleq \mathbf{A}\,\widetilde{\mathbf{s}}(t) \in 
\mathbb{C}^N$, with $\mathbf{A} \triangleq [\mathbf{A}_\text{R}, 
\mathbf{A}_\text{T}] \in \mathbb{C}^{N \times K}$ and 
$\widetilde{\mathbf{s}}(t) \triangleq [\mathbf{s}_\text{R}^\trasp, 
g(t)\,\mathbf{s}_\text{T}^\trasp]^\trasp \in \mathbb{C}^K$.

\subsection{Annihilation of the vector $\mathbf{r}(t)$}
\label{sec:AF-uniform}

We adopt a FRI 
perspective~\cite{pan2016towards,simeoni2020cpgd,pan2018efficient,wang2021two,li2021generic,pan2017frida,guo2022vector} 
to model the STAR-RIS-aided sensing data. The contribution of the $k$-th 
user to the $n$-th entry $r_n(t)$ of $\mathbf{r}(t)$ is 
$\tilde{s}_k(t)\,z_k^{n-1}$, with $z_k \triangleq e^{-j\pi\sin\theta_k}$. 
The superposition of $K$ users across the $N$ metasurface elements 
therefore admits the line-spectrum form
\begin{equation}
r_n(t) = \sum_{k=1}^{K} \tilde{s}_k(t)\,z_k^{n-1}, \quad \text{for $n \in \mathcal{N}$}
\label{eq:line-spectrum}
\end{equation}
where the DOAs are embedded in $z_1,\ldots,z_K$. This structured 
representation reduces DOA estimation to a line-spectrum recovery task 
with $K$ degrees of freedom, enabling gridless recovery via annihilating 
filters and polynomial rooting.

Let $C(z) = \prod_{k=1}^K(1 - z_k^{-1}z) = \sum_{m=0}^K c_m\,z^m$ be 
the $K$-th order annihilating polynomial with roots $z_1,\ldots,z_K$. 
Then, for any shift $\nu \in \{1,\ldots,N-K\}$,
\begin{equation}
\sum_{m=0}^K c_m\,r_{m+\nu}(t) = 
\sum_{k=1}^K \tilde{s}_k(t)\,e^{-j\pi(\nu-1)\sin\theta_k}\,C(z_k) = 0
\label{eq:AF-scalar}
\end{equation}
where the last equality follows from $C(z_k)=0$. In matrix form, this 
reads
\begin{equation}
\mathbf{H}_K\!\left(\mathbf{r}(t)\right)\mathbf{c} = \mathbf{0}_{N-K}, 
\quad \text{for $t \in \mathcal{T}$}
\label{FRI1}
\end{equation}
where $\mathbf{c} \triangleq [c_0,\ldots,c_K]^\trasp \in \mathbb{C}^{K+1}$ 
and the $(N-K)\times(K+1)$ Hankel matrix is given by 
\begin{equation}
\mathbf{H}_K\!\left(\mathbf{r}(t)\right) \triangleq
\begin{bmatrix}
r_1(t) & r_2(t) & \cdots & r_{K+1}(t) \\
r_2(t) & r_3(t) & \cdots & r_{K+2}(t) \\
\vdots & \vdots & \ddots & \vdots \\
r_{N-K}(t) & r_{N-K+1}(t) & \cdots & r_N(t)
\end{bmatrix}.
\label{eq:Hr}
\end{equation}
Stacking \eqref{FRI1} across all $T_\text{s}$ slots, one gives
\begin{equation}
\underbrace{
\begin{bmatrix}
\mathbf{H}_K\!\left(\mathbf{r}(1)\right) \\
\vdots \\
\mathbf{H}_K\!\left(\mathbf{r}(T_\text{s})\right)
\end{bmatrix}
}_{\mathbf{H}_K^\text{stack}(\mathbf{r})}
\mathbf{c} = \mathbf{0}_{(N-K)T_\text{s}}
\label{eq:Hstack}
\end{equation}
where  $\mathbf{R} \eqdef [{\bf{r}}(1), \ldots, {\bf{r}}(T_\text{s})] \in \Cset^{N \times T_\text{s}}$,
with $\mathbf{r} \eqdef \vec(\mathbf{R})$.
Since $\mathbf{H}_K^\text{stack}(\mathbf{r}) \in 
\mathbb{C}^{(N-K)T_\text{s} \times (K+1)}$ has rank $K$, the vector $\mathbf{c}$ 
lies in its null space and can be recovered as the right singular vector 
corresponding to the smallest singular value of 
$\mathbf{H}_K^\text{stack}(\mathbf{r})$ \cite{gonnet2013robust}. 
The DOAs are then obtained from the roots of the polynomial $C(z)$ formed by the entries of $\mathbf{c}$.

\subsection{Proximal gradient descent method}
\label{sec:PGD}

The vector $\mathbf{r}(t)$ is not directly observed at the BS; instead, 
only its compressed projection \eqref{eq:y-final-2} is available. DOA 
recovery therefore requires first estimating a denoised version of 
$\mathbf{r}$ from the measurement vector 
$\mathbf{y} \triangleq [y(1),\ldots,y(T_\text{s})]^\trasp \in \mathbb{C}^{T_\text{s}}$. 
Stacking \eqref{eq:y-final-2} across slots, one gives
\begin{equation}
\mathbf{y} = \boldsymbol{\Phi}\,\mathbf{r} + \mathbf{n}
\label{eq:stacked}
\end{equation}
where $\boldsymbol{\Phi} \triangleq 
\mathrm{blkdiag}[\mathbf{h}^\trasp \boldsymbol{\Phi}_\text{R}(1), \ldots, 
\mathbf{h}^\trasp \boldsymbol{\Phi}_\text{R}(T_\text{s})] \in 
\mathbb{C}^{T_\text{s} \times N T_\text{s}}$ and 
$\mathbf{n} \triangleq [n(1),\ldots,n(T_\text{s})]^\trasp$.
We estimate $\mathbf{r}$ by solving
\begin{equation}
\hat{\mathbf{r}} = \arg\min_{\mathbf{b} \in \mathbb{C}^{NT_\text{s}}} 
\|\mathbf{y} - \boldsymbol{\Phi}\,\mathbf{b}\|_2^2 \quad 
\text{s.t.} \quad 
\mathrm{rank}\!\left(\mathbf{H}_\alpha^\text{stack}(\mathbf{b})\right) \leq K
\label{eq:III-B-origin-rewrite}
\end{equation}
where $\mathbf{H}_\alpha^\text{stack}(\mathbf{b})$ is defined 
as~\eqref{eq:Hstack} with $K$ replaced by $\alpha$ and $\mathbf{r}(t)$ 
by the auxiliary vector $\mathbf{b}(t) \in \Cset^N$, and the rank constraint enforces a $K$-source FRI 
model on the auxiliary variable $\mathbf{b} \eqdef [{\bf{b}}^\trasp(1), \ldots, {\bf{b}}^\trasp(T_\text{s})]^\trasp \in \Cset^{N T_\text{s}}$, ensuring the existence of 
a nontrivial null space from which $\mathbf{c}$ can be recovered. The 
parameter $\alpha$ must satisfy $K \leq \alpha < N$; its optimal choice 
is discussed in Section~\ref{sec:alpha}.

To solve \eqref{eq:III-B-origin-rewrite}, we adopt a 
PGD-based scheme~\cite{ParikhBoyd2014Prox,BeckTeboulle2009FISTA}. 
At iteration $\imath$, the gradient step on the data-fidelity term yields
\begin{equation}
\Delta\mathbf{b}^{(\imath)} = \mathbf{b}^{(\imath)} + 
2\mu_1\,\boldsymbol{\Phi}^\mathrm{H}\!\left(\mathbf{y} - 
\boldsymbol{\Phi}\,\mathbf{b}^{(\imath)}\right)
\label{eq:III-B-gradstep-rewrite}
\end{equation}
where $\mu_1 > 0$ is the step-size (whose admissible range is derived in 
Appendix~\ref{app:mu}). The rank constraint is then enforced via 
alternating projections~\cite{andersson2011alternating} as follows
\begin{equation}
\mathbf{b}^{(\imath+1)} = 
\Pi^{-1}_{\mathcal{H}_\alpha^\text{stack}}\!\left(
\Pi_{\mathcal{H}_\alpha^\text{stack}}\!\left(
\Pi_{\mathcal{R}_K}\!\left(
\mathbf{H}_\alpha^\text{stack}(\Delta\mathbf{b}^{(\imath)})
\right)\right)\right)
\label{eq:III-B-AP-rewrite}
\end{equation}
where $\Pi_{\mathcal{R}_K}(\cdot)$ denotes the rank-$K$ SVD truncation, 
retaining the $K$ largest singular values, 
$\Pi_{\mathcal{H}_\alpha^\text{stack}}(\cdot)$ restores block-Hankel 
structure via block-wise anti-diagonal averaging (Hankelization), and 
$\Pi^{-1}_{\mathcal{H}_\alpha^\text{stack}}(\cdot)$ maps the result back 
to a vector in $\mathbb{C}^{N T_\text{s}}$. The rank-$K$ truncation is not 
exactly stacked-Hankel in general, but is guaranteed to lie close to the 
stacked-Hankel set when $\mathbf{H}_\alpha^\text{stack}(\Delta\mathbf{b}^{(\imath)})$ 
is a small perturbation of a low-rank stacked-Hankel matrix 
(see Appendix~\ref{sec:k-rank_approx_block-Hankel_matrix}).

The iterations are terminated when 
$\|\mathbf{b}^{(\imath+1)} - \mathbf{b}^{(\imath)}\|_2 \leq \varepsilon$ 
or the maximum iteration count $I_\text{max}$ is reached. The resulting 
estimate $\hat{\mathbf{r}}$ is then used to form 
$\mathbf{H}_\alpha^\text{stack}(\hat{\mathbf{r}})$, whose right singular 
vector corresponding to the smallest singular value yields $\hat{\mathbf{c}}$. 
The DOAs are finally recovered as the $K$ roots of $\hat{C}(z)$ lying 
closest to the unit circle. 

The full procedure is summarized in 
Algorithm~\ref{table1}.

\begin{algorithm}[!t]
  \caption{FRI-based DOA estimation for element-wise uniform STAR-RIS}\label{table1}
  \begin{algorithmic}
  \STATE 
  \STATE {\textsc{INPUT}}: $\mathbf{y},N,T_\text{s},K,\alpha,\mu_1, I_\text{max},\varepsilon$
  \STATE {\textsc{OUTPUT}}: ${\hat \theta _1},\cdots,{\hat \theta _K}$
  \STATE $\textbf{Initialize}$ ${\bf{ b}}^{\left(0\right)}$    
  \STATE \textbf{for} $\imath = 1,2, \cdots, I_\text{max}$ \textbf{do}
  \STATE \hspace{0.5cm}$\Delta {{\bf{ b}}^{\left( {\imath} \right)}} \leftarrow {{\bf{b}}^{\left( \imath \right)}} + 2 \, \mu_1 \, 
  {{\bf{\Phi }}^\herm}\left( {{\bf{y}} - {\bf{\Phi }}{{\bf{b}}^{\left( \imath \right)}}} \right)$
  \STATE \hspace{0.5cm}${{{\bf{ b}}}^{\left( {\imath + 1} \right)}} \leftarrow \Pi^{-1}_{\mathcal H_{\alpha}^\text{stack}} \left(\Pi_{\mathcal H_{\alpha}^\text{stack}} \left(
   \Pi_{\mathcal R_K}\left(\mathbf{H}_\alpha^\text{stack}(\Delta\mathbf b^{(\imath)}) \right)\right) \right)$
  \STATE \hspace{0.5cm}$ \textbf{if}$ $\left\| {{{\bf{ b}}^{\left( \imath \right)}} - {{\bf{ b}}^{\left( {\imath + 1} \right)}}} \right\| \le \varepsilon$  
  then $\hat{\mathbf{r}}={\bf{ b}}^{\left( {\imath + 1} \right)}$ \textbf{break}
  \STATE \hspace{0.5cm}\textbf{else} 
  \STATE  \hspace{0.5cm} $\hat{\mathbf{r}}={\bf{ b}}^{\left( {\imath + 1} \right)}$
  \STATE \hspace{0.5cm} $\imath = \imath + 1$
  \STATE \textbf{end for} $\imath$
  \STATE ${\bf{ U}},{\bf{S}},{\bf{V}} \leftarrow {\text{SVD}}\left( \mathbf{H}_\alpha^\text{stack}(\hat{\mathbf{r}})\right)$
  \STATE $\hat{C}\left( z \right) \leftarrow \hat{\mathbf{c}}={\bf{V}}\left( {:,\alpha  + 1} \right)$
  \STATE $\textbf{Keep}$ the $K$ roots $\{{\hat z}_k\}_{k=1}^K$ of $\hat{C}(z)$ closest to unit circle
  \STATE $ {\hat \theta _k} = -\arcsin \left( {{{\arg \left( {{\hat z}_k} \right)} \mathord{\left/
  {\vphantom {{\arg \left( {{z_k}} \right)} \pi }} \right.
  \kern-\nulldelimiterspace} \pi }} \right)$ for $k \in \{1,2,\ldots, K\}$
  \end{algorithmic}
  \end{algorithm}

\section{Full-Space DOA Estimation Under Nonuniform Energy-Splitting 
STAR-RIS Parameters}
\label{sec:Nonuniform}

Although the STAR-RIS parameters are programmable, enforcing identical 
settings across all elements can be overly restrictive. In practice, 
improving communication throughput and sensing accuracy often requires 
additional degrees of freedom in the metasurface design, particularly in 
the allocation of reflection/transmission power. Under narrowband 
operation with identical unit cells, the reflection--transmission phase 
offset is primarily determined by the element design and is approximately 
invariant across the array~\cite{Bao2021}. We therefore consider a more 
general regime in which the phase difference is common across the 
metasurface, while the ES coefficients vary across elements.

By virtue of \eqref{eq:j} and \eqref{eq:p}, the matrix 
$\widetilde{\mathbf{A}}(t)$ partitions as
\be
\widetilde{\mathbf{A}}(t) = \left[\mathbf{A}_\text{R},\; 
\mathbf{G}(t) \mathbf{A}_\text{T}\right]
\label{eq:Atildepart-nonunif}
\ee
with $\mathbf{G}(t) \triangleq \pm j\,\diag[p_1(t),\ldots,p_N(t)]$.
Substituting \eqref{eq:Atildepart-nonunif} into \eqref{eq:y-final} gives
\be
y(t) = \boldsymbol{\psi}^\trasp(t)\,\mathbf{x} + n(t)
\label{eq:y-final-3}
\ee
where $\boldsymbol{\psi}(t) \triangleq [\mathbf{I}_N, 
\mathbf{G}(t)]^\trasp\boldsymbol{\Phi}_\text{R}(t)\,\mathbf{h} \in 
\mathbb{C}^{2N}$ and we have defined $\mathbf{x} \triangleq [\mathbf{x}_\text{R}^\trasp, 
\mathbf{x}_\text{T}^\trasp]^\trasp \in \mathbb{C}^{2N}$ (cf.\ 
\eqref{eq:x}). Collecting all slots, one has
\be
\mathbf{y} = \boldsymbol{\Psi}^\trasp\,\mathbf{x} + \mathbf{n}
\label{eq:stacked-nonunif}
\ee
with $\boldsymbol{\Psi} \triangleq [\boldsymbol{\psi}(1),\ldots,
\boldsymbol{\psi}(T_\text{s})] \in \mathbb{C}^{2N \times T_\text{s}}$.

\subsection{Annihilation of the vector $\mathbf{x}$}
\label{sec:AF-nonuniform}

Although the element-wise ES coefficients vary, the reflected and 
transmitted components still admit parallel FRI-type representations. 
Specifically, along the same lines of Subsection~\ref{sec:AF-uniform}, 
the following AF equations hold:
\be
\mathbf{H}_{K_i}(\mathbf{x}_i)\,\mathbf{c}_i = \mathbf{0}_{N-K_i}\:,
\quad \text{for } i \in \{\text{R},\text{T}\}
\label{FRI2}
\ee
where $\mathbf{H}_{K_i}(\mathbf{x}_i) \in 
\mathbb{C}^{(N-K_i)\times(K_i+1)}$ is the Hankel matrix obtained from 
\eqref{eq:Hr} by replacing $K$ and $\mathbf{r}(t)$ with $K_i$ and 
$\mathbf{x}_i$ respectively, and $\mathbf{c}_i \in \mathbb{C}^{K_i+1}$ 
collects the annihilation coefficients. Since each entry of $\mathbf{x}_i$ 
is a superposition of $K_i$ complex exponentials, 
$\mathrm{rank}(\mathbf{H}_{K_i}(\mathbf{x}_i)) = K_i$, so $\mathbf{c}_i$ 
is the right singular vector of $\mathbf{H}_{K_i}(\mathbf{x}_i)$ 
corresponding to its smallest singular value. The angles are then 
recovered from the roots of the polynomial $C_i(z)$ with coefficients 
$\mathbf{c}_i$. In the proposed algorithm, the two constraints 
\eqref{FRI2} are not imposed independently but enforced jointly via a 
single horizontally concatenated lifting operator, as described next.

\subsection{Proximal gradient descent method}
\label{sec:PGD-nonunif}

DOA recovery requires first estimating a denoised version of the latent 
FRI vector $\mathbf{x}$ from $\mathbf{y}$. We introduce the auxiliary 
vector $\boldsymbol{\beta} \triangleq [\boldsymbol{\beta}_\text{R}^\trasp, 
\boldsymbol{\beta}_\text{T}^\trasp]^\trasp \in \mathbb{C}^{2N}$ and 
define, for $i \in \{\text{R},\text{T}\}$, the Hankel matrix 
$\mathbf{H}_\alpha(\boldsymbol{\beta}_i) \in 
\mathbb{C}^{(N-\alpha)\times(\alpha+1)}$ as in \eqref{eq:Hr} with $K$ 
and $\mathbf{r}(t)$ replaced by $\alpha$ and $\boldsymbol{\beta}_i$. The 
\emph{paired lifting} is 
\begin{equation}
\mathbf{H}_\alpha^\mathrm{pair}(\boldsymbol{\beta}) \triangleq 
\left[\mathbf{H}_\alpha(\boldsymbol{\beta}_\mathrm{R})\;\;
\mathbf{H}_\alpha(\boldsymbol{\beta}_\mathrm{T})\right]
\in \mathbb{C}^{(N-\alpha)\times 2(\alpha+1)}
\label{eq:Hpair-def}
\end{equation}
which has rank at most $K = K_\text{R} + K_\text{T}$ under the FRI 
generative model. We estimate $\mathbf{x}$ by solving
\begin{equation}
\hat{\mathbf{x}} = \arg\min_{\boldsymbol{\beta} \in \mathbb{C}^{2N}} 
\|\mathbf{y} - \boldsymbol{\Psi}^\trasp\boldsymbol{\beta}\|_2^2 \quad 
\text{s.t.} \quad 
\mathrm{rank}\!\left(\mathbf{H}_\alpha^\text{pair}(\boldsymbol{\beta})
\right) \leq K
\label{eq:PGD_pair_form}
\end{equation}
where the rank constraint enforces the joint FRI structure on both 
subspaces simultaneously. The PGD solver follows the same steps as in 
Subsection~\ref{sec:PGD}: at each iteration $\imath$, a gradient step on 
the data-fidelity term gives
\begin{equation}
\Delta\boldsymbol{\beta}^{(\imath)} = \boldsymbol{\beta}^{(\imath)} + 
2\mu_2\,\boldsymbol{\Psi}^*\!\left(\mathbf{y} - 
\boldsymbol{\Psi}^\trasp\boldsymbol{\beta}^{(\imath)}\right)
\label{eq:gradstep-nonunif}
\end{equation}
followed by alternating projections to enforce the paired low-rank 
Hankel structure:
\begin{equation}
\boldsymbol{\beta}^{(\imath+1)} = 
\Pi^{-1}_{\mathcal{H}_\alpha^\text{pair}}\!\left(
\Pi_{\mathcal{H}_\alpha^\text{pair}}\!\left(
\Pi_{\mathcal{R}_K}\!\left(
\mathbf{H}_\alpha^\text{pair}(\Delta\boldsymbol{\beta}^{(\imath)})
\right)\right)\right)
\label{eq:AP-nonunif}
\end{equation}
where $\Pi_{\mathcal{R}_K}(\cdot)$ performs rank-$K$ SVD truncation as 
in \eqref{eq:III-B-AP-rewrite}, 
$\Pi_{\mathcal{H}_\alpha^\text{pair}}(\cdot)$ restores the paired Hankel 
structure via anti-diagonal averaging on each block, and 
$\Pi^{-1}_{\mathcal{H}_\alpha^\text{pair}}(\cdot)$ maps the result back 
to two vectors in $\mathbb{C}^N$. When 
$\mathbf{H}_\alpha^\text{pair}(\Delta\boldsymbol{\beta}^{(\imath)})$ is a 
small perturbation of an ideal low-rank paired Hankel matrix, the rank-$K$ 
SVD approximation is guaranteed to lie close to the paired Hankel set in 
Frobenius norm, by the same argument as 
Appendix~\ref{sec:k-rank_approx_block-Hankel_matrix}. 

The overall procedure is summarized in Algorithm~\ref{table2}, where $\mu_2 > 0$ is 
the PGD step-size.

\begin{algorithm}[!t]
  \caption{FRI-based DOA estimation for element-wise nonuniform STAR-RIS}\label{table2}
  \begin{algorithmic}
  \STATE 
  \STATE {\textsc{INPUT}}: $\mathbf{y},N,T_\text{s},K, K_\text{T},K_\text{R},\alpha,\mu_2, I_\text{max},\varepsilon$
  \STATE {\textsc{OUTPUT}}: ${\hat \theta _1},\cdots,{\hat \theta_K}$
  \STATE $\textbf{Initialize}$ $\boldsymbol{\beta}^{\left(0\right)}$    
  \STATE \textbf{for} $\imath = 1,2, \cdots, I_\text{max}$ \textbf{do}
   \STATE \hspace{0.5cm} $\Delta \boldsymbol{\beta}^{\left( {\imath} \right)} \leftarrow \boldsymbol{\beta}^{\left( \imath \right)} + 2 \, \mu_2 \, 
 {{\bf{\Psi }}^*}\left( {{\bf{y}} - {\bf{\Psi }}^\trasp {\boldsymbol{\beta}^{\left( \imath \right)}}} \right)$
\STATE \hspace{0.5cm} ${\boldsymbol{\beta}^{\left( {\imath + 1} \right)}} \leftarrow \Pi^{-1}_{\mathcal H_{\alpha}^\text{pair}} \left(\Pi_{\mathcal H_{\alpha}^\text{pair}} 
\left(\Pi_{\mathcal R_K}\left(\mathbf{H}_\alpha^\text{pair}(\Delta\boldsymbol{\beta}^{(\imath)}) \right)\right) \right)$
  \STATE \hspace{0.5cm}$ \textbf{if}$ $\| \boldsymbol{\beta}^{\left( \imath \right)} - \boldsymbol{\beta}^{\left( {\imath + 1} \right)} \| \le \varepsilon$  
  then $\hat{\mathbf{x}}=\boldsymbol{\beta}^{\left( {\imath + 1} \right)}$ \textbf{break}
  \STATE \hspace{0.5cm}\textbf{else} 
  \STATE  \hspace{0.5cm} $\hat{\mathbf{x}}=\boldsymbol{\beta}^{\left( {\imath + 1} \right)}$
  \STATE \hspace{0.5cm} $\imath = \imath + 1$
\STATE \textbf{end for} $\imath$
  \STATE ${\bf{ U}}_i,{\bf{S}}_i,{\bf{V}}_i \leftarrow {\text{SVD}}\left( \mathbf{H}_\alpha(\hat{\mathbf{x}}_i)\right)$ for $i\in\{\text{T},\text{R}\}$
  \STATE $\hat{C}_i(z)$ $\leftarrow$ $\hat{\mathbf{c}}_i={\bf{V}}_i\left( {:,\alpha  + 1} \right)$ for $i\in\{\text{T},\text{R}\}$
  \STATE $\textbf{Keep}$ the $K_\text{R}$ roots $\{{\hat z}_k\}_{k \in \mathcal{K}_\text{R}}$ of $\hat{C}_\text{R}(z)$ 
  and the $K_\text{T}$ roots $\{{\hat z}_k\}_{k \in \mathcal{K}_\text{T}}$ of $\hat{C}_\text{T}(z)$ closest to unit circle
  \STATE $ {\hat \theta _k} = -\arcsin \left( {{{\arg \left( {{\hat z}_k} \right)} \mathord{\left/
  {\vphantom {{\arg \left( {{z_k}} \right)} \pi }} \right.
  \kern-\nulldelimiterspace} \pi }} \right)$ for $k \in \{1,2,\ldots, K\}$
 \end{algorithmic}
  \end{algorithm}

\section{Parameter Setting and Performance Analysis}
\label{sec:Perform}

The performance of the proposed algorithms depends critically on two 
design parameters: the PGD step-size and the Hankel lifting dimension. 
These choices jointly govern convergence behavior, numerical stability, 
the effectiveness of the low-rank projection, the resolvability of 
closely spaced angles, and computational complexity.

\subsection{Choice of the step-size}
\label{sec:stepsize}

The step-size $\mu_1$ of Algorithm~\ref{table1} governs the 
proximal-gradient update and directly impacts whether the alternating 
projections yield an effective descent direction. 
A sufficient local stability condition is
\begin{equation}
\frac{1}{2\lambda_\text{max}}\!\left(1 - \frac{1}{\sqrt{\alpha+1}}\right)
< \mu_1 <
\frac{1}{2\lambda_\text{max}}\!\left(1 + \frac{1}{\sqrt{\alpha+1}}\right)
\label{conver1}
\end{equation}
where $\lambda_\text{max} \eqdef \sigma_{\max}^2(\boldsymbol{\Phi})$ is the 
largest eigenvalue of $\boldsymbol{\Phi}^\herm\boldsymbol{\Phi}$ 
(see Appendix~\ref{app:mu}). By the same contraction 
argument, the step-size $\mu_2$ of Algorithm~\ref{table2} satisfies an 
analogous condition with $\lambda_\text{max}$ replaced by 
$\sigma_{\max}^2(\boldsymbol{\Psi})$, where $\boldsymbol{\Psi}$ is 
defined in \eqref{eq:stacked-nonunif}.

\subsection{Choice of the lifting parameter $\alpha$}
\label{sec:alpha}

The parameter $\alpha$ controls the dimension of the lifted Hankel 
matrices, thus influencing convergence efficiency, the maximum number of 
resolvable sources, and noise robustness. The feasibility condition on 
$\alpha$ differs between the two algorithms because the rank constraint 
is imposed on different lifted objects.

In Algorithm~\ref{table1}, the FRI model order is encoded in each 
slot-wise Hankel block $\mathbf{H}_\alpha(\mathbf{b}(t)) \in 
\mathbb{C}^{(N-\alpha)\times(\alpha+1)}$, and vertical stacking increases 
only the number of rows without enlarging the column dimension. Under the 
constraint $\mathrm{rank}(\mathbf{H}_\alpha(\mathbf{b}(t))) \leq K$, the 
null space has dimension at least $(\alpha+1)-K$, provided $\alpha \geq 
K$. A larger null-space dimension improves noise robustness by providing 
more redundant algebraic constraints, while the rank-$K$ truncation 
requires sufficient degrees of freedom in both dimensions, i.e.,
\begin{equation}
K \leq \min\{\alpha+1,\; N-\alpha\}.
\label{eq:K1}
\end{equation}
We adopt $\alpha = \alpha_1 \triangleq \lfloor N/2 \rfloor$ for 
Algorithm~\ref{table1}.

In Algorithm~\ref{table2}, the rank prior is enforced on the paired 
lifting $\mathbf{H}_\alpha^\text{pair}(\boldsymbol{\beta}) = 
[\mathbf{H}_\alpha(\boldsymbol{\beta}_\text{R}),\, 
\mathbf{H}_\alpha(\boldsymbol{\beta}_\text{T})]$, whose column dimension 
is $2(\alpha+1)$. The feasibility condition becomes
\begin{equation}
K \leq \min\{2(\alpha+1),\; N-\alpha\}
\label{eq:K2}
\end{equation}
and we set $\alpha = \alpha_2 \triangleq \lfloor N/3 \rfloor$. The 
enlarged column dimension in \eqref{eq:K2} allows a higher admissible 
model order than \eqref{eq:K1} for a fixed aperture size, which 
explains the superior performance of Algorithm~\ref{table2} over 
Algorithm~\ref{table1}, as confirmed by the numerical results reported in 
Section~\ref{sec:results}.

\subsection{Computational complexity}

The dominant cost of both algorithms is the truncated SVD implementing 
the rank-$K$ projection. The gradient step and Hankelization operations 
require only matrix-vector products and block-wise averaging, incurring 
$\mathcal{O}(T_\text{s} N^2)$ per iteration, which is lower than the SVD cost 
for moderate-to-large $N$.

For Algorithm~\ref{table1}, the stacked-Hankel lifting yields an SVD of 
size $(N-\alpha)T_\text{s} \times (\alpha+1)$; with $\alpha \approx N/2$ this 
scales as $\mathcal{O}(T_\text{s}\lceil N/2\rceil^3)$ per iteration. For 
Algorithm~\ref{table2}, the paired lifting produces a near-square matrix 
of size $\lceil 2N/3\rceil \times \lceil 2N/3\rceil$ under $\alpha \approx 
N/3$, with per-iteration cost $\mathcal{O}(\lceil 2N/3\rceil^3)$. In the 
large-array regime the cubic dependence on $N$ dominates, while the 
linear scaling with $T_\text{s}$ has a milder impact. Both methods converge in a 
limited number of iterations, keeping the overall computational burden 
moderate.

\subsection{Ziv-Zakai bound}
\label{sec:ZZB}

This subsection derives a coupled full-space ZZB for the proposed 
STAR-RIS-assisted sensing model, building directly on the unified 
observation model \eqref{eq:y-final} so that the 
reflection/transmission coupling is preserved throughout.

Let $\boldsymbol{\theta} \triangleq [\theta_1,\ldots,\theta_K]^\trasp = 
[\boldsymbol{\theta}_\text{R}^\trasp, \boldsymbol{\theta}_\text{T}^\trasp]^\trasp 
\in \mathbb{R}^K$ collect the full-space DOAs, conditioned on the known 
vector $\mathbf{s}$. The stacked observation satisfies 
$\mathbf{y} \sim \mathcal{CN}(\boldsymbol{\mu}(\boldsymbol{\theta};\mathbf{s}),\, 
\sigma_n^2 \, \mathbf{I}_{T_\text{s}})$ with mean
\begin{equation}
\mu_t(\boldsymbol{\theta};\mathbf{s}) =
\mathbf{h}^\trasp\boldsymbol{\Phi}_\text{R}(t)\widetilde{\mathbf{A}}(t) \, \mathbf{s}
= \sum_{k=1}^{K} s_k\,\mathbf{h}^\trasp\boldsymbol{\Phi}_\text{R}(t) \, 
\widetilde{\mathbf{a}}(t;\theta_k).
\label{eq:zzb_mean_t}
\end{equation}
Defining $\boldsymbol{\gamma}_k(\theta_k) \triangleq 
[\gamma_k(1;\theta_k),\ldots,\gamma_k(T_\text{s};\theta_k)]^\trasp$ with 
$\gamma_k(t;\theta_k) \triangleq \mathbf{h}^\trasp\boldsymbol{\Phi}_\text{R}(t) 
\widetilde{\mathbf{a}}(t;\theta_k)$, the mean vector decomposes as
\begin{equation}
\boldsymbol{\mu}(\boldsymbol{\theta};\mathbf{s}) =
\boldsymbol{\mu}_\text{R}(\boldsymbol{\theta}_\text{R};\mathbf{s}_\text{R}) +
\boldsymbol{\mu}_\text{T}(\boldsymbol{\theta}_\text{T};\mathbf{s}_\text{T})
\label{eq:zzb_mu_split}
\end{equation}
where $\boldsymbol{\mu}_i \triangleq \sum_{k\in\mathcal{K}_i} 
s_k\,\boldsymbol{\gamma}_k(\theta_k)$ for $i \in \{\text{R},\text{T}\}$.

For the binary hypothesis test $\mathcal{H}_0:\boldsymbol{\theta}$ 
vs.\ $\mathcal{H}_1:\boldsymbol{\theta}+\boldsymbol{\delta}$, the 
mean difference is
\begin{equation}
\Delta\boldsymbol{\mu}(\boldsymbol{\theta},\boldsymbol{\delta}) \triangleq
\boldsymbol{\mu}(\boldsymbol{\theta}+\boldsymbol{\delta};\mathbf{s}) -
\boldsymbol{\mu}(\boldsymbol{\theta};\mathbf{s}) =
\Delta\boldsymbol{\mu}_\text{R} + \Delta\boldsymbol{\mu}_\text{T}
\label{eq:zzb_delta_mu}
\end{equation}
and the pairwise detection distance decomposes as
\be
d_\text{STAR}^2(\boldsymbol{\theta},\boldsymbol{\delta})
\triangleq \frac{1}{\sigma_n^2}
\|\Delta\boldsymbol{\mu}(\boldsymbol{\theta},\boldsymbol{\delta})\|_2^2
= d_\text{R}^2 + d_\text{T}^2 + d_\text{RT}^2
\label{eq:zzb_distance_decomp}
\ee
where $d_i^2 \triangleq \sigma_n^{-2} \, \|\Delta\boldsymbol{\mu}_i\|_2^2$ 
for $i \in \{\text{R},\text{T}\}$ and, additionally,  
$d_\text{RT}^2 \triangleq 2 \, \sigma_n^{-2} \,  
\Re\{\Delta\boldsymbol{\mu}_\text{R}^\herm \, \Delta\boldsymbol{\mu}_\text{T}\}$. 
The cross-term $d_\text{RT}^2$ is induced by simultaneous transmission 
and reflection and vanishes under subspace-wise separated or 
switching-based sensing.

Let $\Theta_0 \triangleq [-\pi/2, \pi/2]^{K_\text{R}} \times 
[-\pi/2, \pi/2]^{K_\text{T}}$ denote the prior support of the full-space 
DOA vector $\boldsymbol{\theta}$, and let us denote with 
$h_{\max}(\mathbf{u}) \triangleq 
\sup\{h \geq 0 : \Theta(h,\mathbf{u}) \neq \emptyset\}$ the maximum 
displacement in direction $\mathbf{u}$ compatible with the support. 
Under equal prior probabilities, the minimum pairwise error probability (PEP) 
is $P_{\min} = \mathcal{Q}(d_\text{STAR}/\sqrt{2})$. Following the 
standard vector-parameter ZZB framework~\cite{Zhang2023Ziv}, for any unit-norm 
direction $\mathbf{u} \in \mathbb{R}^K$, one gets
\begin{equation}
\mathbf{u}^\trasp\mathbf{C}_{\hat{\boldsymbol{\theta}}}\mathbf{u} \geq
\frac{1}{2}\int_0^{h_{\max}(\mathbf{u})} h\,
\mathcal{V}\!\left\{\bar{P}_{\min}(h,\mathbf{u})\right\} {\rm d}h
\label{eq:zzb_directional}
\end{equation}
where $\mathbf{C}_{\hat{\boldsymbol{\theta}}}$ is the estimation error 
covariance matrix, $\mathcal{V}\{\cdot\}$ is the valley-filling operator, and
\be
\bar P_{\min}(h,\mathbf u) \eqdef
\frac{1}{|\Theta(h,\mathbf u)|} \int_{\Theta(h,\mathbf u)} P_{\min}(\boldsymbol{\theta},\boldsymbol{\theta}+h\mathbf u) d\boldsymbol{\theta}
\label{eq:zzb_pbar}
\ee
is the prior-averaged PEP over the valid displacement region
$\Theta(h,\mathbf{u}) \triangleq \{\boldsymbol{\theta}: 
\boldsymbol{\theta}\in\Theta_0,\, 
\boldsymbol{\theta}+h\mathbf{u}\in\Theta_0\}$. 
Averaging over canonical directions $\mathbf{e}_m$ gives a lower bound 
on the mean square error (MSE) of any unbiased estimator of 
$\boldsymbol{\theta}$:
\begin{multline}
\mathrm{MSE}_\text{full} \triangleq \frac{1}{K} \, \mathrm{Tr}
\{\mathbf{C}_{\hat{\boldsymbol{\theta}}}\}  \\ \geq
\frac{1}{2K}\sum_{m=1}^{K}\int_0^{h_{\max}(\mathbf{e}_m)}
h\,\mathcal{V}\!\left\{\bar{P}_{\min}(h,\mathbf{e}_m)\right\} {\rm d}h.
\label{eq:zzb_mse}
\end{multline}
The full-space FIM is $\mathbf{J}_{\boldsymbol{\theta}}^\text{STAR} = 
(2/\sigma_n^2) \, \Re\{\mathbf{G}^\herm\mathbf{G}\}$, where 
we have introduced the matrix $\mathbf{G} \triangleq [\mathbf{g}_1,\ldots,\mathbf{g}_K]$ with
\begin{equation}
\mathbf{g}_k \triangleq \frac{\partial\boldsymbol{\mu}}{\partial\theta_k}
= \left[s_k\,\mathbf{h}^\trasp\boldsymbol{\Phi}_\text{R}(t)
\left(\mathbf{j}_k(t)\odot\mathbf{p}_k(t)\odot
\dot{\mathbf{a}}(\theta_k)\right)\right]_{t=1}^{T_\text{s}}
\label{eq:zzb_gk}
\end{equation}
and $\dot{\mathbf{a}}(\theta_k) \triangleq \partial\mathbf{a}(\theta_k)/
\partial\theta_k$. By partitioning $\mathbf{G} = [\mathbf{G}_\text{R}, 
\mathbf{G}_\text{T}]$, one gives the block FIM
\begin{equation}
\mathbf{J}_{\boldsymbol{\theta}}^\text{STAR} =
\begin{bmatrix}
\mathbf{J}_\text{RR} & \mathbf{J}_\text{RT} \\
\mathbf{J}_\text{TR} & \mathbf{J}_\text{TT}
\end{bmatrix}
\label{eq:zzb_fim_block}
\end{equation}
where $\mathbf{J}_{ij} = (2/\sigma_n^2) \, \Re\{\mathbf{G}_i^\herm 
\mathbf{G}_j\}$ for $i,j \in \{\text{R},\text{T}\}$. The off-diagonal 
block $\mathbf{J}_\text{RT}$ is generally nonzero, capturing 
cross-subspace information coupling under simultaneous sensing.

To obtain a computable closed-form bound, we adopt a blockwise ordered 
prior $\Theta_0^\text{ord} \triangleq \Theta_\text{R}^\text{ord} \times 
\Theta_\text{T}^\text{ord}$ and the permutation-invariant error metric
\begin{equation}
d^2(\hat{\boldsymbol{\theta}},\boldsymbol{\theta}) \triangleq
\min_{\pi_\text{R}\in\mathfrak{S}_{K_\text{R}}}
\|\hat{\boldsymbol{\theta}}_\text{R} - 
\pi_\text{R}(\boldsymbol{\theta}_\text{R})\|_2^2
+\min_{\pi_\text{T}\in\mathfrak{S}_{K_\text{T}}}
\|\hat{\boldsymbol{\theta}}_\text{T} - 
\pi_\text{T}(\boldsymbol{\theta}_\text{T})\|_2^2
\label{eq:zzb_block_perm_mse}
\end{equation}
where $\mathfrak{S}_{K_i}$ is the symmetric group of degree $K_i$. The 
a priori performance bound (APB) for each subspace is
\begin{equation}
\mathrm{APB}_i^\text{ord} =
\frac{K_i\zeta_i^2}{(K_i+1)^2(K_i+2)}\:,
\quad \text{for $i \in \{\text{R},\text{T}\}$}
\label{eq:zzb_apb_subspaces}
\end{equation}
where $\zeta_i$ is the effective angular width of subspace $i$, and the 
full-space APB is given by
\begin{equation}
\mathrm{APB}_\text{full}^\text{ord} \triangleq
\frac{K_\text{R}\,\mathrm{APB}_\text{R}^\text{ord} +
K_\text{T}\,\mathrm{APB}_\text{T}^\text{ord}}{K}.
\label{eq:zzb_apb_full}
\end{equation}

The effective coupled SNR is defined as
\begin{equation}
\eta_\text{eff}^\text{STAR} \triangleq
\frac{\|\boldsymbol{\mu}_\text{R}\|_2^2 + \|\boldsymbol{\mu}_\text{T}\|_2^2
+ 2\Re\{\boldsymbol{\mu}_\text{R}^\herm \, \boldsymbol{\mu}_\text{T}\}}
{NKT_\text{s}\sigma_n^2}
\label{eq:zzb_eta_eff}
\end{equation}
where the cross term captures the coupling induced by simultaneous 
transmission and reflection. The large-error probability is approximated as
\begin{multline}
P_L^\text{STAR} =
\exp\!\left(KT_\text{s}\!\left[\ln\!\frac{4(1+N\eta_\text{eff}^\text{STAR})}
{(2+N\eta_\text{eff}^\text{STAR})^2}
+ \left(\frac{N\eta_\text{eff}^\text{STAR}}
{2+N\eta_\text{eff}^\text{STAR}}\right)^{\!2}\right]\right)
\\
\times\;\mathcal{Q}\!\left(\sqrt{2KT_\text{s}}\,
\frac{N\eta_\text{eff}^\text{STAR}}{2+N\eta_\text{eff}^\text{STAR}}\right)
\label{eq:zzb_PL_star}
\end{multline}
and the transition parameter is
\begin{equation}
\tilde{u}_\text{STAR} \approx
\min\!\left\{KT_\text{s}\!\left(\frac{N\eta_\text{eff}^\text{STAR}}
{2+N\eta_\text{eff}^\text{STAR}}\right)^{\!2},\;
\frac{K^2\zeta_\text{eff}^2}
{8\,\mathbf{1}_K^\trasp
(\mathbf{J}_{\boldsymbol{\theta}}^\text{STAR})^{-1}\mathbf{1}_K}
\right\}
\label{eq:zzb_u_star}
\end{equation}
with $\zeta_\text{eff}^2 \triangleq (K_\text{R}\zeta_\text{R}^2 + 
K_\text{T}\zeta_\text{T}^2)/K$. The coupled full-space ZZB is then given by
\begin{equation}
\mathrm{MSE}_\text{full} \geq
2P_L^\text{STAR}\,\mathrm{APB}_\text{full}^\text{ord}
+ \Gamma_{\frac{3}{2}}(\tilde{u}_\text{STAR})\,
\frac{\mathrm{Tr}\{(\mathbf{J}_{\boldsymbol{\theta}}^\text{STAR})^{-1}\}}{K}
\label{eq:zzb_final_closed}
\end{equation}
where the first term dominates at low SNR and reflects the ordered prior 
geometry, while the second term approaches the coupled CRB at high SNR.

\section{Numerical Results}
\label{sec:results}

This section assesses the feasibility, robustness, and angle-estimation 
accuracy of the proposed methods through numerical simulations. We 
consider a STAR-RIS-assisted uplink with a half-wavelength ULA having 
$N=16$ elements, where all users transmit narrowband signals over the 
same carrier frequency. Unless otherwise stated, we set $K=4$ users with 
identical SNR, with two users located in the RS and two in the TS. We 
investigate two STAR-RIS settings: \emph{Scenario~1} (uniform 
parameters), where $\beta_{\text{R},n}=\beta_{\text{T},n}=\sqrt{2}/2$ 
for all $n$; and \emph{Scenario~2} (nonuniform parameters), where 
$\beta_{\text{R},n}^2$ is uniformly distributed in $[0.2,0.8]$, subject 
to \eqref{eq:power}, and $\beta_{\text{T},n}$ is determined accordingly. 
To keep the figure legends concise, ``M1, Scen.~1'' refers to the 
proposed Method~1 evaluated under Scenario~1, and similarly for the 
other combinations.

\subsection{Experiment~1: AF spectrum and root validation}

To validate the recovered AF solutions and the corresponding polynomial 
roots, we adopt the semi-space angle representation. The RS angles are 
set to $\boldsymbol{\theta}_\text{R} = [-12.23^\circ,\, 39.19^\circ]$ 
and the TS angles to $\boldsymbol{\theta}_\text{T} = [-47.34^\circ,\, 
15.57^\circ]$, with $\text{SNR} = 15$~dB. We scan $\theta \in 
[-60^\circ, 60^\circ]$ and plot the AF spectra: $C(\theta)$ corresponds 
to Algorithm~\ref{table1}, whereas $\{C(\theta_\text{R}), 
C(\theta_\text{T})\}$ correspond to the RS/TS subspace equations in 
Algorithm~\ref{table2}.

\begin{figure}[!t]
\centering
\subfloat[Scenario 1]{%
  \includegraphics[width=0.48\linewidth]{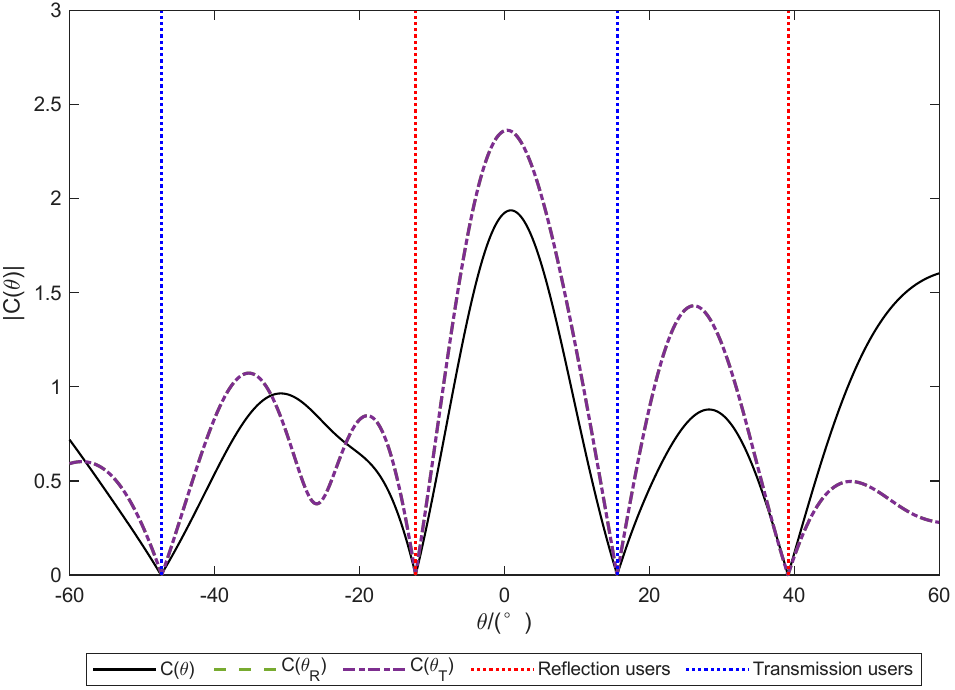}%
  \label{Spectrum1}%
}\hfill
\subfloat[Scenario 2]{%
  \includegraphics[width=0.48\linewidth]{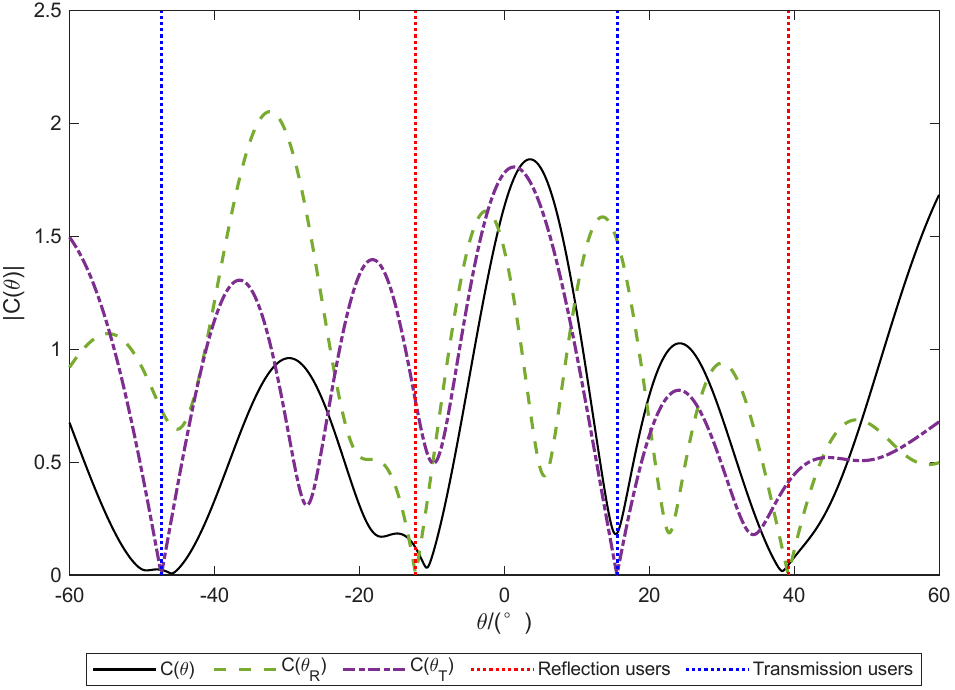}%
  \label{Spectrum2}%
}
\caption{Spectrum of the AF equations under different scenarios.}
\label{fig:Spectrum}
\end{figure}

Figs.~\ref{Spectrum1} and \ref{Spectrum2} show that, in Scenario~1, all 
spectra exhibit pronounced nulls at the true angles. Moreover, 
$C(\theta_\text{R}) \approx C(\theta_\text{T})$, which is consistent 
with the induced rank-one mapping in the uniform-parameter regime: the 
reconstructed RS/TS submatrices are highly correlated and therefore yield 
nearly identical null spaces. In Scenario~2, this rank-one coupling is 
violated. As a result, the single-equation spectrum $C(\theta)$ produced 
by Algorithm~\ref{table1} no longer attains exact zeros and its minima 
become biased, whereas Algorithm~\ref{table2} maintains accurate nulls by 
enforcing the two subspace AF constraints.

\begin{figure}[t]
{\includegraphics[width=\linewidth]{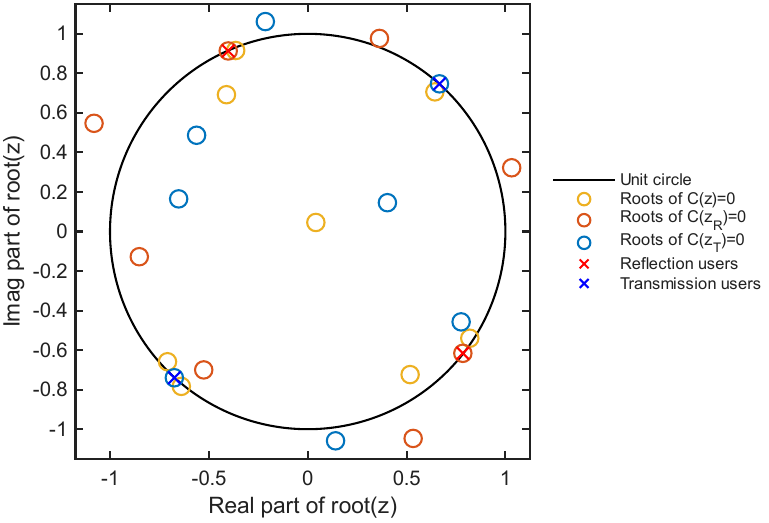}}
\caption{Root distribution of AF equations (Scenario~2).}
\label{fig:root}
\end{figure}

As illustrated in Fig.~\ref{fig:root}, the AF polynomial typically yields 
more roots than the number of sources; spurious candidates are removed by 
retaining only the $K$ roots lying closest to the unit circle. Under 
Scenario~2, Algorithm~\ref{table1} fails to preserve the correct root 
locations, while Algorithm~\ref{table2} consistently retains the roots 
associated with the true angles.

\subsection{Experiment~2: Estimation accuracy and success probability}

We evaluate the full-space angle estimation capability of the proposed 
approaches, focusing on the angular region in which reliable estimation 
is attainable. The full-space representation is adopted throughout. 
Performance is quantified in terms of (i) estimation accuracy and (ii) 
success probability. A trial is declared successful if the maximum 
absolute error satisfies $\max_k |\hat{\theta}_k - \theta_k| \leq 
5^\circ$, and the success probability is defined as $p = L_s/L$, where 
$L_s$ denotes the number of successful runs among $L$ Monte Carlo 
trials. The accuracy metric is the RMSE computed over successful trials 
only:
\begin{equation}
\text{RMSE} = \sqrt{\frac{1}{KL_s} \sum_{l=1}^{L_s} \sum_{k=1}^{K}
\left(\hat{\theta}_{k,l} - \theta_k\right)^2}.
\label{eq:RMSE}
\end{equation}
For the RMSE evaluation, the estimated angles are matched to the true 
angles by solving the optimal assignment problem within each subspace 
separately, i.e., the RS estimates are matched to RS true angles and the 
TS estimates to TS true angles, using the minimum-cost permutation.

\begin{figure}[!t]
\centering
\subfloat[RMSE performance of full-space DOA estimation]{%
  \includegraphics[width=0.48\linewidth]{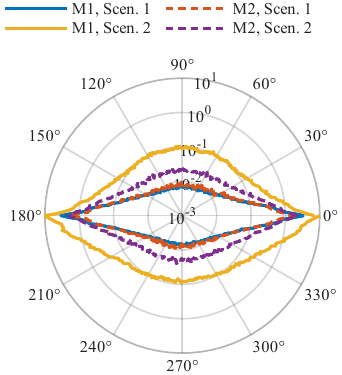}%
  \label{fig:fullRMSE}%
}\hfill
\subfloat[Success probability of full-space DOA estimation]{%
  \includegraphics[width=0.48\linewidth]{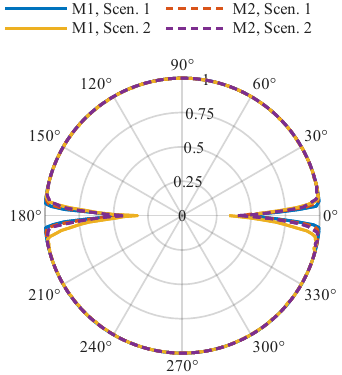}%
  \label{fig:Success}%
}
\caption{Evaluation of full-space DOA estimation performance.}
\label{fig:Evaluation}
\end{figure}

We consider a two-user setting with one user located in the RS and the 
other in the TS. One user's angle sweeps the full domain with a $1^\circ$ 
step, while the other user's angle is randomly drawn in the opposite 
subspace. For each swept angle, $L = 1000$ Monte Carlo trials are 
conducted. The resulting RMSE and success-probability curves are reported 
in Fig.~\ref{fig:Evaluation} (M1 and M2 stand for Method~1 and 
Method~2, respectively).

As shown in Fig.~\ref{fig:fullRMSE}, both Algorithm~\ref{table1} and 
Algorithm~\ref{table2} exhibit pronounced performance degradation as the 
impinging direction approaches the coplanar (grazing) case, which is 
consistent with classical array-based DOA behavior. 
Fig.~\ref{fig:Success} further indicates that within $[-80^\circ, 
80^\circ]$ (semi-space representation), the proposed algorithms maintain 
success probability close to one and achieve an accuracy on the order of 
$0.1^\circ$. Overall, the results confirm the feasibility of full-space 
estimation in STAR-RIS systems. To ensure consistently high success 
probability, we restrict user angles to $[-60^\circ, 60^\circ]$ 
(semi-space representation) in the subsequent experiments.

\subsection{Experiment~3: Convergence behavior}

\begin{figure}[t]
{\includegraphics[width=\linewidth]{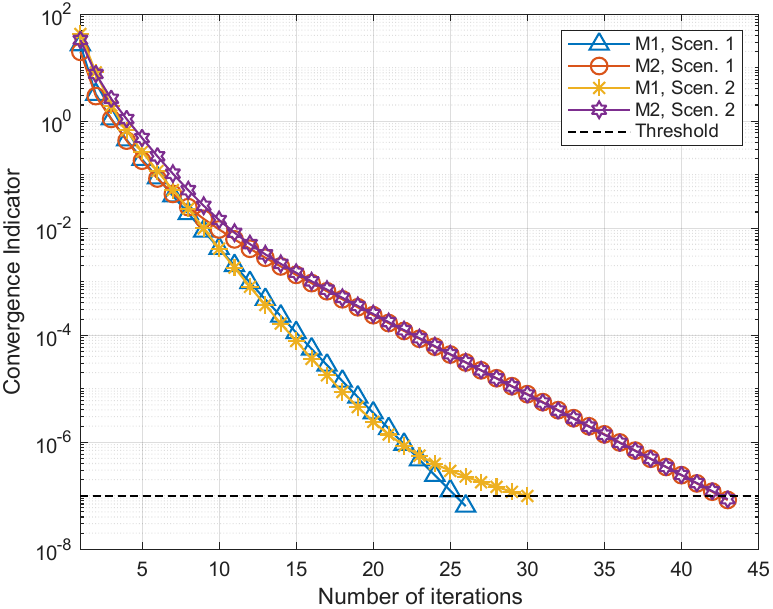}}
\caption{Convergence behavior of the proposed algorithms.}
\label{fig:convergence}
\end{figure}

Convergence within a finite number of iterations is essential for both 
reliability and computational efficiency. Lack of convergence indicates 
that the PGD iterates do not approach a stationary point, whereas slow 
convergence translates into excessive runtime. This experiment 
investigates whether the proposed step-size condition in \eqref{conver1} 
leads to stable and rapid convergence. Fig.~\ref{fig:convergence} depicts 
the averaged convergence behavior over $1000$ Monte Carlo trials. The 
stopping tolerance is set to $\varepsilon = 10^{-7}$, and the step-sizes 
are selected strictly within the admissible interval in \eqref{conver1}.

Both Algorithm~\ref{table1} and Algorithm~\ref{table2} converge within 
$45$ iterations, and the error decreases to the order of $10^{-5}$ 
within approximately $30$ iterations. In Scenario~1, the two algorithms 
exhibit nearly identical exponential decay during the first $20$ 
iterations. Afterwards, Algorithm~\ref{table1} continues a steady descent 
and reaches the stopping threshold earlier, whereas 
Algorithm~\ref{table2} shows a slower tail. This behavior is mainly due 
to the joint block-matrix reconstruction in Algorithm~\ref{table2}: as 
the iterates approach the structured low-rank manifold, the remaining 
residual becomes dominated by weak noise components, for which the 
rank-$K$ truncation on the concatenated block matrix yields smaller 
incremental reductions.

In Scenario~2, the two algorithms follow similar convergence 
trajectories. However, convergence in this regime only indicates that the 
PGD iterations have reached an approximate stationary point of the 
adopted model; the mismatch induced by nonuniform STAR-RIS coefficients 
may still bias the recovered AF solutions. Overall, these results confirm 
the practical viability of the proposed solvers and show that the 
prescribed step-size selection provides stable convergence.

\subsection{Experiment~4: RMSE versus SNR}

\begin{figure}[!t]
\centering
\subfloat[Scenario 1]{%
  \includegraphics[width=0.48\linewidth]{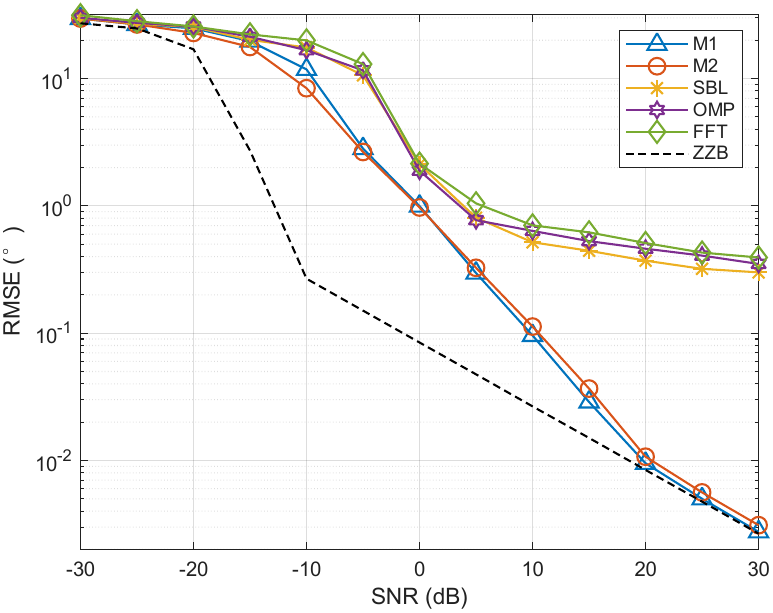}%
  \label{fig:snr1}%
}\hfill
\subfloat[Scenario 2]{%
  \includegraphics[width=0.48\linewidth]{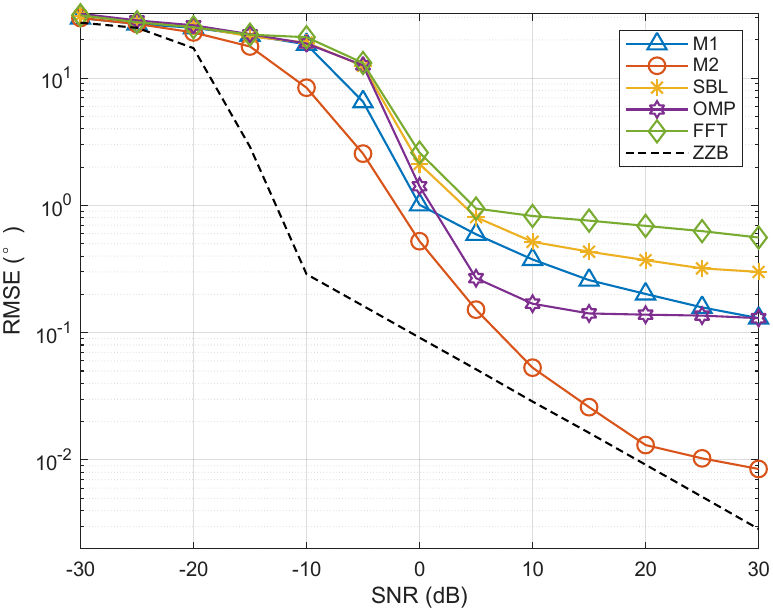}%
  \label{fig:snr2}%
}
\caption{RMSE versus SNR under different scenarios.}
\label{fig:SNR}
\end{figure}

We compare the proposed approaches with three representative baselines, 
namely FFT-based beam scanning, orthogonal matching pursuit (OMP)~\cite{TroppGilbert2007}, and 
SBL~\cite{Gerstoft2016SBL}, all implemented with an angular grid 
resolution of $0.01^\circ$. All baselines estimate the RS and TS angles 
separately. To assess noise robustness, we sweep the SNR from $-30$~dB 
to $30$~dB in $5$~dB increments. The resulting RMSE curves under 
Scenario~1 and Scenario~2 are reported in Fig.~\ref{fig:SNR}.

As shown in Fig.~\ref{fig:snr1}, under Scenario~1 both proposed methods 
exhibit nearly identical trends and consistently outperform FFT/OMP/SBL 
over the entire SNR range. In this regime, the element-wise uniform 
STAR-RIS setting preserves the induced rank-one full-space mapping, so 
both M1 and M2 remain well matched to the underlying multichannel FRI 
structure. Consequently, their RMSE decreases steadily with SNR and 
closely approaches the ZZB after the bound leaves its low-SNR \emph{a 
priori} region and enters the CRB-like decay regime. A closer inspection 
reveals a mild crossover between the two proposed methods: M2 is slightly 
better at low SNR, whereas M1 becomes marginally better at high SNR. 
This trend is physically meaningful. At low SNR, M2 benefits from the 
enlarged paired lifting, whose stronger structural redundancy improves 
low-rank denoising. At high SNR, the thermal-noise contribution is 
already weak and the remaining error is mainly determined by 
finite-dimensional projection and root-extraction bias. Since the 
rank-one coupling is exactly preserved in Scenario~1, the simpler 
single-structure formulation of M1 is already sufficient and may incur a 
slightly smaller implementation bias than the larger paired lifting in M2.

The situation changes significantly in Scenario~2, as shown in 
Fig.~\ref{fig:snr2}. The nonuniform ES coefficients violate the rank-one 
coupling underlying the uniform-parameter model, so the single AF 
relation exploited by M1 is no longer sufficient to provide accurate root 
estimates, which explains its clear degradation. By contrast, M2 remains 
substantially more robust because it explicitly preserves the two-subspace 
structure through the paired lifting. Nevertheless, unlike Scenario~1, a 
mild high-SNR floor is observed for M2. This residual gap reflects the 
fact that in Scenario~2 the paired-Hankel structure is only approximate 
rather than exact, as supported by Appendix~\ref{sec:k-rank_approx_block-Hankel_matrix}. 
Hence, once the thermal-noise effect becomes negligible, the remaining 
error is dominated by residual structural mismatch, finite-dimensional 
low-rank projection, and root-selection bias, which together produce the 
observed small plateau.

A similar interpretation applies to the baseline methods. The angular 
search grid for FFT/OMP/SBL is already as fine as $0.01^\circ$, which is 
sufficiently dense for practical grid-based estimation. Hence, the 
observed high-SNR floor cannot be attributed to grid quantization alone. 
Even at $\text{SNR} = 30$~dB, the RMSE of these baselines does not reach 
the grid precision and remains in a plateau region. This indicates that 
their dominant error source at high SNR is not insufficient angular 
resolution, but rather the residual structured mismatch induced by the 
STAR-RIS full-space coupled observation model. Since these baselines 
estimate the RS and TS angles separately, the contribution from the 
opposite subspace enters the estimation process as an unmodeled 
structured interference term. Consequently, neither increasing the SNR 
nor refining the angular grid is sufficient to remove the resulting bias, 
which explains the persistent high-SNR floor of FFT/OMP/SBL and further 
highlights the importance of explicitly exploiting the 
reflection--transmission coupling.

Overall, these results clarify the distinct operating regimes of M1 and 
M2. The former is highly competitive when the STAR-RIS follows the ideal 
uniform regime, whereas the latter becomes essential once the STAR-RIS 
parameters depart from that idealized structure.

\subsection{Experiment~5: Computational cost}

This experiment benchmarks the computational efficiency of the proposed 
algorithms against the FFT-, OMP-, and SBL-based baselines. 
Table~\ref{tab1} reports the average runtime per Monte Carlo run (in 
seconds) measured in MATLAB, for both Scenario~1 and Scenario~2.

\begin{table}[t]
\centering
\caption{Running time of different methods.}
\label{tab1}
\begin{tabular}{ccc}
\toprule
\textbf{Method} & \textbf{Scenario 1} & \textbf{Scenario 2} \\
\midrule
FFT      & $8.9\times 10^{-3}$~s & $9.1\times 10^{-3}$~s \\
OMP      & $7.4\times 10^{-3}$~s & $8.3\times 10^{-3}$~s \\
SBL      & $3.2\times 10^{-2}$~s & $4.3\times 10^{-2}$~s \\
Method~1 & $8.1\times 10^{-3}$~s & $8.9\times 10^{-3}$~s \\
Method~2 & $1.4\times 10^{-2}$~s & $2.4\times 10^{-2}$~s \\
\bottomrule
\end{tabular}
\end{table}

As shown in Table~\ref{tab1}, FFT- and OMP-based methods exhibit the 
lowest runtime due to their lightweight grid-based processing. 
Algorithm~\ref{table1} attains a runtime of the same order, indicating 
that the structured PGD--AP solver incurs only limited additional 
overhead. Algorithm~\ref{table2} is more computationally demanding, 
consistent with the complexity analysis in 
Section~\ref{sec:Perform}: it relies on a higher-dimensional lifted 
matrix and therefore requires a costlier truncated-SVD projection at each 
iteration. The runtime gap becomes more evident in Scenario~2, where the 
nonuniform-parameter setting typically leads to slower convergence and 
less favorable numerical conditioning.

\subsection{Experiment~6: Effect of STAR-RIS size}

\begin{figure}[t]
{\includegraphics[width=\linewidth]{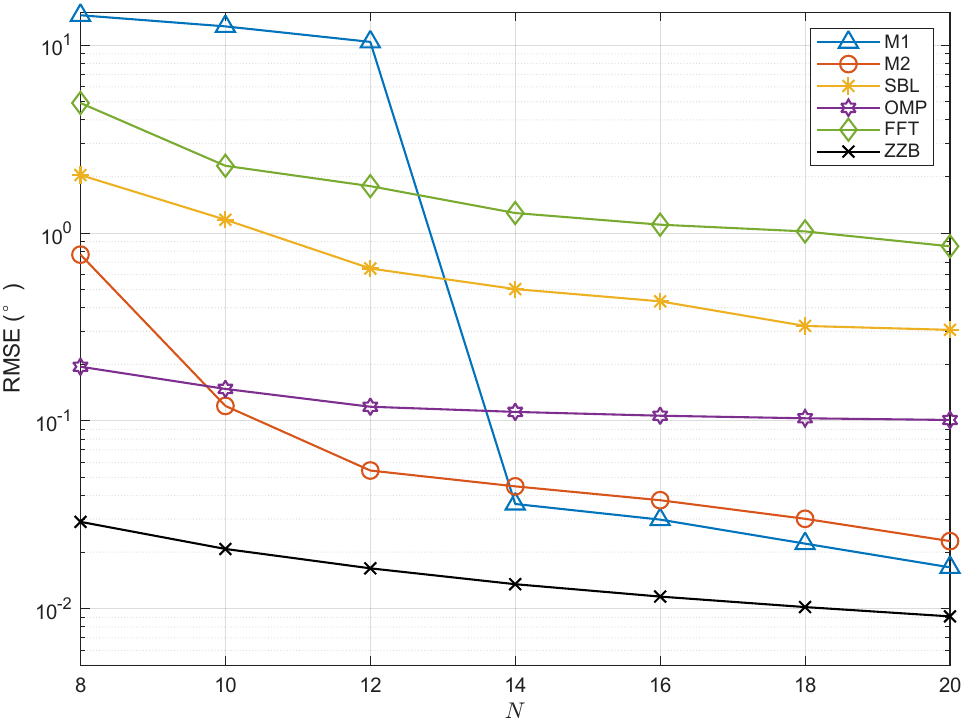}}
\caption{RMSE versus the number of elements (Scenario~1).}
\label{fig:elements}
\end{figure}

This experiment examines the impact of the STAR-RIS aperture on 
estimation accuracy by varying the number of metasurface elements from 
$N=8$ to $N=20$ in steps of $2$ under Scenario~1. The corresponding 
RMSE results are reported in Fig.~\ref{fig:elements}.

It can be seen that the proposed algorithms exhibit a stronger dependence 
on $N$ than the three baselines. Once a minimum aperture requirement is 
met (e.g., $N > K$), FFT/OMP/SBL show only a mild improvement with 
increasing $N$. In contrast, the proposed FRI/Hankel-based solvers 
benefit directly from a larger lifting dimension, which strengthens the 
structured low-rank constraint, improves denoising, and stabilizes the 
subsequent root-extraction step.

For small apertures, Algorithm~\ref{table1} may approach (or violate) 
the feasibility limits of the rank-$K$ truncation due to insufficient 
lifted degrees of freedom, resulting in unstable estimation. As $N$ 
increases, its performance improves rapidly and approaches the best 
attainable accuracy. Algorithm~\ref{table2} remains applicable over the 
entire range of $N$ thanks to the paired formulation, but its RMSE 
degrades in the small-$N$ regime because reduced Hankel dimensions weaken 
the SVD-based noise suppression and make the annihilation-based recovery 
more sensitive to perturbations.

Overall, these results highlight an inherent trade-off of the proposed 
structured lifting: smaller STAR-RIS apertures limit the effectiveness of 
the low-rank model, whereas larger apertures enhance denoising capability 
and angular resolvability.

\section{Conclusion}
\label{sec:conclusions}

This paper investigated gridless full-space DOA estimation in 
STAR-RIS-assisted ISAC systems by exploiting the FRI
structure induced by the coupled reflection/transmission mechanism of the 
metasurface. Two recovery schemes were developed: an efficient algorithm 
tailored to the element-wise uniform energy-splitting regime, and a more 
general paired-lifting method that remains applicable under element-wise 
nonuniform settings. Both schemes combine structured low-rank denoising 
via PGD with alternating projections onto a 
block-Hankel matrix set, followed by DOA retrieval through 
AF root finding. 
Sufficient local stability conditions on the PGD step-size
and the lifting parameter were established.
A Ziv-Zakai bound was also derived for the coupled full-space sensing 
model, accounting explicitly for the cross-subspace coupling induced by 
simultaneous transmission and reflection, and used to benchmark the 
absolute performance limits of the proposed methods.

Numerical results confirmed reliable full-space coverage and consistent 
gains over grid-based baselines across the full SNR range, while 
highlighting two inherent limitations: performance degradation near 
grazing angles, which is common to all array-based DOA methods, and a 
mild residual error floor under strong nonuniformity, which stems from 
the approximate rather than exact paired-Hankel structure in that regime. 
The proposed framework is particularly attractive for
low-cost STAR-RIS sensing architectures with a single RF
chain, where covariance-based methods become difficult to
apply due to the lack of repeated stationary snapshots.

The proposed framework assumes a single-path model and slot-invariant 
complex gains, which are appropriate for narrowband pilot-aided sensing 
intervals; extending the approach to multipath environments and 
time-varying channels represents a natural direction for future work.

\appendices

\section{Approximate Stacked-Hankel Structure After SVD Truncation}
\label{sec:k-rank_approx_block-Hankel_matrix}

Let $\mathcal{H}_\alpha^\text{stack}$ denote the linear subspace of 
matrices having the stacked-Hankel structure induced by the lifting 
operator $\mathbf{H}_\alpha^\text{stack}(\cdot)$. Given 
$\mathbf{Q}_\alpha^\text{stack} \in \mathbb{C}^{(N-\alpha)T_\text{s} \times 
(\alpha+1)}$, Algorithm~\ref{table1} forms its rank-$K$ truncation via 
SVD as
\begin{equation}
\mathbf{Q}_\alpha^\text{stack} = \mathbf{U}\boldsymbol{\Sigma}\mathbf{V}^\herm
\quad\Rightarrow\quad
\mathbf{Q}_K^\text{stack} \triangleq \mathbf{U}\boldsymbol{\Sigma}_K\mathbf{V}^\herm
\label{eq:app_svdtrunc}
\end{equation}
where $\boldsymbol{\Sigma}_K$ retains only the $K$ largest singular 
values. Since $\mathcal{H}_\alpha^\text{stack}$ is independent of the 
rank constraint, $\mathbf{Q}_K^\text{stack}$ is not guaranteed to be 
stacked Hankel. Nevertheless, under the FRI generative model it is 
\emph{approximately} stacked Hankel, in the sense that its Frobenius 
distance to $\mathcal{H}_\alpha^\text{stack}$ is small.

To show this, assume that $\mathbf{Q}_\alpha^\text{stack}$ can be 
decomposed as
\begin{equation}
\mathbf{Q}_\alpha^\text{stack} = \mathbf{H}^\text{stack} + \mathbf{E}
\label{eq:app_decomp}
\end{equation}
where $\mathbf{H}^\text{stack} \in \mathcal{H}_\alpha^\text{stack}$ is 
an ideal stacked-Hankel matrix generated by a $K$-sparse line-spectrum 
sequence, and $\mathbf{E}$ collects noise and modeling errors whose 
entries are i.i.d.\ zero-mean complex circular Gaussian with variance 
$\sigma_e^2$. The FRI property implies $\mathrm{rank}(\mathbf{H}^\text{stack}) 
\leq K$, so $\sigma_{K+1}(\mathbf{H}^\text{stack}) = 0$.

Let $\mathrm{dist}(\mathbf{X}, \mathcal{H}_\alpha^\text{stack}) \triangleq 
\min_{\mathbf{Z} \in \mathcal{H}_\alpha^\text{stack}} \|\mathbf{X} - 
\mathbf{Z}\|_F$ denote the Frobenius distance to the stacked-Hankel 
subspace. Since $\mathbf{H}^\text{stack} \in \mathcal{H}_\alpha^\text{stack}$, 
we have
\begin{multline}
\mathrm{dist}(\mathbf{Q}_K^\text{stack}, \mathcal{H}_\alpha^\text{stack})
\leq \|\mathbf{Q}_K^\text{stack} - \mathbf{H}^\text{stack}\|_F \\
\leq \|\mathbf{Q}_K^\text{stack} - \mathbf{Q}_\alpha^\text{stack}\|_F 
+ \|\mathbf{E}\|_F.
\label{eq:app_dist1}
\end{multline}
We now bound each term separately. By the Eckart--Young theorem 
\cite{EckartYoung1936}, the best rank-$K$ approximation satisfies
\begin{equation}
\|\mathbf{Q}_\alpha^\text{stack} - \mathbf{Q}_K^\text{stack}\|_F^2
= \sum_{i > K} \sigma_i^2(\mathbf{Q}_\alpha^\text{stack}).
\label{eq:app_EY}
\end{equation}
By Weyl's inequality \cite{Horn}, each singular value of 
$\mathbf{Q}_\alpha^\text{stack} = \mathbf{H}^\text{stack} + \mathbf{E}$ 
satisfies
\begin{equation}
\sigma_{K+1}(\mathbf{Q}_\alpha^\text{stack})
\leq \sigma_{K+1}(\mathbf{H}^\text{stack}) + \|\mathbf{E}\|_2
= \|\mathbf{E}\|_2
\label{eq:app_weyl}
\end{equation}
where the equality uses $\sigma_{K+1}(\mathbf{H}^\text{stack}) = 0$ and 
$\|\mathbf{E}\|_2$ denotes the spectral norm of $\mathbf{E}$. Since 
$\sigma_i(\mathbf{Q}_\alpha^\text{stack}) \leq \|\mathbf{E}\|_2$ for all 
$i > K$, substituting into \eqref{eq:app_EY} gives
\begin{equation}
\|\mathbf{Q}_\alpha^\text{stack} - \mathbf{Q}_K^\text{stack}\|_F
\leq \|\mathbf{Q}_\alpha^\text{stack} - \mathbf{H}^\text{stack}\|_F
= \|\mathbf{E}\|_F
\label{eq:app_EY_simple}
\end{equation}
where the last equality follows from \eqref{eq:app_decomp}. Substituting 
\eqref{eq:app_EY_simple} into \eqref{eq:app_dist1} yields
\begin{equation}
\mathrm{dist}(\mathbf{Q}_K^\text{stack}, \mathcal{H}_\alpha^\text{stack})
\leq \|\mathbf{Q}_K^\text{stack} - \mathbf{Q}_\alpha^\text{stack}\|_F
+ \|\mathbf{E}\|_F \leq 2\|\mathbf{E}\|_F.
\label{eq:app_finalbound}
\end{equation}

Therefore, the rank-$K$ SVD truncation $\mathbf{Q}_K^\text{stack}$ is 
guaranteed to lie within a Frobenius-norm ball of radius $2\|\mathbf{E}\|_F$ 
around the stacked-Hankel subspace. Since $\mathbf{H}^\text{stack}$ is 
formed by vertically stacking Hankel sub-blocks, the bound 
\eqref{eq:app_finalbound} immediately implies that each reconstructed 
sub-block $\mathbf{Q}_{K,t}$ of $\mathbf{Q}_K^\text{stack}$ is itself 
close to the Hankel set $\mathcal{H}_\alpha$:
\begin{equation}
\mathrm{dist}(\mathbf{Q}_{K,t}, \mathcal{H}_\alpha) 
\leq \mathrm{dist}(\mathbf{Q}_K^\text{stack}, \mathcal{H}_\alpha^\text{stack})
\leq 2\|\mathbf{E}\|_F\:,
\quad \forall\, t \in \mathcal{T}.
\label{eq:app_blockbound}
\end{equation}

Since the noise $\mathbf{E}$ is modeled as circularly symmetric complex 
Gaussian, the squared Frobenius norm $\|\mathbf{E}\|_F^2 / \sigma_e^2$ 
follows a chi-squared distribution with $2M$ degrees of freedom, where 
$M \triangleq (N-\alpha)T_\text{s}(\alpha+1)$ is the number of complex entries. 
Invoking the Laurent-Massart concentration inequality for chi-squared 
random variables \cite{Vershynin2018HDP}, for any $\delta \in (0,1)$,
\begin{equation}
\|\mathbf{E}\|_F \leq \sigma_e\!\left[\sqrt{M} + 
\sqrt{2\ln(\delta^{-1})}\right]
\label{eq:app_concentration}
\end{equation}
with probability at least $1-\delta$. Combining \eqref{eq:app_finalbound} 
and \eqref{eq:app_concentration}, the rank-$K$ truncation satisfies
\begin{equation}
\mathrm{dist}(\mathbf{Q}_K^\text{stack}, \mathcal{H}_\alpha^\text{stack})
\leq 2\sigma_e\!\left[\sqrt{M} + \sqrt{2\ln(\delta^{-1})}\right]
\label{eq:app_final_prob}
\end{equation}
with probability at least $1-\delta$, thereby confirming that the approximation 
error vanishes as 
$\sigma_e \to 0$.

\section{Local Contraction Analysis and Step-Size Selection}
\label{app:mu}

We provide a sufficient local stability condition for the AP-PGD 
iteration in Algorithm~\ref{table1} by showing that the composite update 
defines a contraction mapping in a neighborhood of the ideal low-rank 
solution. The analysis is carried out under the assumption that the 
dominant rank-$K$ singular subspace remains unchanged within this 
neighborhood. Algorithm~\ref{table2} follows by applying the same 
argument to the corresponding block operator.

Define the gradient step associated with the quadratic data-fidelity term 
as
\begin{equation}
P_{\mu_1}(\mathbf{a}) \triangleq \mathbf{a} + 
2\mu_1\,\boldsymbol{\Phi}^\herm(\mathbf{y} - \boldsymbol{\Phi}\,\mathbf{a})
\label{eq:app_Pmu}
\end{equation}
and the composite AP-induced update as
\begin{equation}
U_{\mu_1}(\mathbf{a}) \triangleq
\Pi^{-1}_{\mathcal{H}_\alpha^\text{stack}}\!\left(
\Pi_{\mathcal{H}_\alpha^\text{stack}}\!\left(
\widetilde{\Pi}_{\mathcal{R}_K}\!\left(
\mathbf{H}_\alpha^\text{stack}(P_{\mu_1}(\mathbf{a}))
\right)\right)\right)
\label{eq:app_Umu}
\end{equation}
where $\widetilde{\Pi}_{\mathcal{R}_K}(\cdot)$ denotes the fixed-subspace 
projection defined in Step~3 below.

\smallskip
\noindent\textbf{Step 1: Gradient step.}
Since $\boldsymbol{\Phi} \in \mathbb{C}^{T_\text{s} \times NT_\text{s}}$ has 
$T_\text{s} \ll NT_\text{s}$, the matrix $\boldsymbol{\Phi}^\herm\boldsymbol{\Phi}$ is 
rank deficient with $\lambda_\text{min} = 0$. Consequently, $P_{\mu_1}$ 
acts as the identity on $\ker(\boldsymbol{\Phi})$, and the standard 
spectral bound $\rho_{\mu_1} \geq 1$ does not yield a contraction 
globally. However, since the data-fidelity term depends on $\mathbf{b}$ 
only through $\boldsymbol{\Phi}\mathbf{b}$, the gradient step introduces 
no update along $\ker(\boldsymbol{\Phi})$, i.e., 
\begin{equation}
P_{\mu_1}(\mathbf{a}) - \mathbf{a} = 
2\mu_1\boldsymbol{\Phi}^\herm(\mathbf{y} - \boldsymbol{\Phi}\mathbf{a}) 
\in \mathcal{R}(\boldsymbol{\Phi}^\herm)
\label{eq:app_Pmu_range}
\end{equation}
so the contraction needs to be established only on the row space 
$\mathcal{R}(\boldsymbol{\Phi}^\herm)$. On this subspace, the eigenvalues 
of $\boldsymbol{\Phi}^\herm\boldsymbol{\Phi}$ range from 
$\lambda_\text{min}^+ \triangleq \sigma_\text{min}^{+2}(\boldsymbol{\Phi}) 
> 0$ (the smallest nonzero eigenvalue) to 
$\lambda_\text{max} \triangleq \sigma_\text{max}^2(\boldsymbol{\Phi})$, and,  
for any $\mathbf{a}_1, \mathbf{a}_2 \in \mathbb{C}^{NT_\text{s}}$, 
\begin{multline}
\|P_{\mu_1}(\mathbf{a}_1) - P_{\mu_1}(\mathbf{a}_2)\|_2 
= \|(\mathbf{I}_{NT_\text{s}} - 2\mu_1\boldsymbol{\Phi}^\herm\boldsymbol{\Phi})
(\mathbf{a}_1 - \mathbf{a}_2)\|_2 
\\ \leq \rho_{\mu_1}\,\|\mathbf{a}_1 - \mathbf{a}_2\|_2
\label{eq:app_P_lip}
\end{multline}
where
\begin{equation}
\rho_{\mu_1} \triangleq \max\left\{
|1 - 2\mu_1\lambda_\text{min}^+|,\;
|1 - 2\mu_1\lambda_\text{max}|
\right\}
\label{eq:app_rho}
\end{equation}
is the spectral norm of $\mathbf{I}_{NT_\text{s}} - 
2\mu_1\boldsymbol{\Phi}^\herm\boldsymbol{\Phi}$ restricted to 
$\mathcal{R}(\boldsymbol{\Phi}^\herm)$. The null-space component does 
not affect the data-fidelity residual and is implicitly regularized by 
the rank-$K$ Hankel constraint.

\smallskip
\noindent\textbf{Step 2: Hankel lifting.}
The lifting operator satisfies
\begin{equation}
\|\mathbf{H}_\alpha^\text{stack}(\mathbf{a}_1) - 
\mathbf{H}_\alpha^\text{stack}(\mathbf{a}_2)\|_F
\leq \sqrt{\alpha+1}\,\|\mathbf{a}_1 - \mathbf{a}_2\|_2
\label{eq:app_H_lip}
\end{equation}
because each entry of $\mathbf{a}_1 - \mathbf{a}_2$ appears at most 
$(\alpha+1)$ times in $\mathbf{H}_\alpha^\text{stack}(\cdot)$, so the 
Frobenius norm aggregates at most $(\alpha+1)$ copies of each squared 
magnitude.

\smallskip
\noindent\textbf{Step 3: Fixed-subspace rank-$K$ projection.}
The rank-$K$ truncation $\Pi_{\mathcal{R}_K}(\cdot)$ via SVD is 
\emph{not} non-expansive in general. To recover a contraction, we 
restrict to a local perturbation regime in which the iterates satisfy
\be
\|\mathbf{H}_\alpha^\text{stack}(\Delta\mathbf{b}^{(\imath)}) - 
\mathbf{H}^\text{stack}\|_F \leq \delta
\ee
for some small $\delta > 0$, so that the dominant rank-$K$ signal 
subspace does not change across iterations. Let 
$\mathbf{H}^\text{stack} = \mathbf{U}_\star\boldsymbol{\Sigma}_\star
\mathbf{V}_\star^\herm$ be the compact SVD of the ideal stacked-Hankel 
matrix, with $\mathbf{U}_\star \in \mathbb{C}^{(N-\alpha)T_\text{s} \times K}$ 
and $\mathbf{V}_\star \in \mathbb{C}^{(\alpha+1)\times K}$ having 
orthonormal columns, and define the orthogonal projectors 
$\mathbf{P}_U \triangleq \mathbf{U}_\star\mathbf{U}_\star^\herm$ and 
$\mathbf{P}_V \triangleq \mathbf{V}_\star\mathbf{V}_\star^\herm$. 
The fixed-subspace projection is given by 
\begin{equation}
\widetilde{\Pi}_{\mathcal{R}_K}(\mathbf{A}) \triangleq 
\mathbf{P}_U\,\mathbf{A}\,\mathbf{P}_V
\label{eq:app_fixed_proj}
\end{equation}
which retains the component of $\mathbf{A}$ in the dominant rank-$K$ 
signal subspace and discards the noise-subspace complement. 
By the Davis-Kahan theorem~\cite{Horn}, when 
\be
\|\mathbf{H}_\alpha^\text{stack}(\Delta\mathbf{b}^{(\imath)}) - 
\mathbf{H}^\text{stack}\|_F \leq \delta
\ee 
the principal subspaces of 
$\mathbf{H}_\alpha^\text{stack}(\Delta\mathbf{b}^{(\imath)})$ and 
$\mathbf{H}^\text{stack}$ differ by at most 
$\mathcal{O}(\delta/\Delta\sigma)$ in the sine of the principal angles, 
where $\Delta\sigma$ is the gap between the $K$-th and $(K+1)$-th 
singular values of $\mathbf{H}^\text{stack}$. Consequently, 
$\Pi_{\mathcal{R}_K}$ and $\widetilde{\Pi}_{\mathcal{R}_K}$ agree up to 
a perturbation of order $\mathcal{O}(\delta/\Delta\sigma)$, and the 
approximation improves as $\delta/\Delta\sigma \to 0$.

We now show that $\widetilde{\Pi}_{\mathcal{R}_K}(\cdot)$ is 
non-expansive in Frobenius norm. For any $\mathbf{A}_1, \mathbf{A}_2 
\in \mathbb{C}^{(N-\alpha)T_\text{s} \times (\alpha+1)}$:
\begin{align}
\|\widetilde{\Pi}_{\mathcal{R}_K}(\mathbf{A}_1) - 
\widetilde{\Pi}_{\mathcal{R}_K}(\mathbf{A}_2)\|_F
&= \|\mathbf{P}_U(\mathbf{A}_1 - \mathbf{A}_2)\mathbf{P}_V\|_F 
\nonumber\\
&\leq \|\mathbf{P}_U\|_2\,\|\mathbf{A}_1 - \mathbf{A}_2\|_F\,
\|\mathbf{P}_V\|_2 \nonumber\\
&= \|\mathbf{A}_1 - \mathbf{A}_2\|_F
\label{eq:app_svd_nonexp_revised}
\end{align}
where the last equality uses $\|\mathbf{P}_U\|_2 = \|\mathbf{P}_V\|_2 = 
1$, since $\mathbf{P}_U$ and $\mathbf{P}_V$ are orthogonal projectors.

\smallskip
\noindent\textbf{Step 4: Inverse Hankelization.}
The inverse Hankelization operator is non-expansive from Frobenius to 
Euclidean norm:
\begin{equation}
\|\Pi^{-1}_{\mathcal{H}_\alpha^\text{stack}}(\mathbf{A}_1) - 
\Pi^{-1}_{\mathcal{H}_\alpha^\text{stack}}(\mathbf{A}_2)\|_2
\leq \|\mathbf{A}_1 - \mathbf{A}_2\|_F.
\label{eq:app_Hinv_nonexp}
\end{equation}

\smallskip
\noindent\textbf{Contraction and step-size condition.}
Combining \eqref{eq:app_P_lip}--\eqref{eq:app_Hinv_nonexp} under the 
local perturbation assumption, one yields
\begin{equation}
\|U_{\mu_1}(\mathbf{a}_1) - U_{\mu_1}(\mathbf{a}_2)\|_2
\leq \sqrt{\alpha+1}\,\rho_{\mu_1}\,\|\mathbf{a}_1 - \mathbf{a}_2\|_2.
\label{eq:app_contraction}
\end{equation}
For $U_{\mu_1}$ to be a contraction on $\mathcal{R}(\boldsymbol{\Phi}^\herm)$, 
we require $\sqrt{\alpha+1}\,\rho_{\mu_1} < 1$, i.e.,
\begin{equation}
\max\left\{|1 - 2\mu_1\lambda_\text{min}^+|,\;
|1 - 2\mu_1\lambda_\text{max}|\right\} < \frac{1}{\sqrt{\alpha+1}}.
\label{eq:app_contraction_cond}
\end{equation}
This amounts to requiring simultaneously
$|1 - 2\mu_1\lambda_\text{min}^+| < 1/\sqrt{\alpha+1}$ and
$|1 - 2\mu_1\lambda_\text{max}| < 1/\sqrt{\alpha+1}$,
which yield respectively the intervals
\begin{align}
\frac{1}{2\lambda_\text{min}^+}\!\left(1 - \frac{1}{\sqrt{\alpha+1}}\right)
&< \mu_1 <
\frac{1}{2\lambda_\text{min}^+}\!\left(1 + \frac{1}{\sqrt{\alpha+1}}\right)
\label{eq:app_mu_lmin}\\
\frac{1}{2\lambda_\text{max}}\!\left(1 - \frac{1}{\sqrt{\alpha+1}}\right)
&< \mu_1 <
\frac{1}{2\lambda_\text{max}}\!\left(1 + \frac{1}{\sqrt{\alpha+1}}\right).
\label{eq:app_mu_lmax}
\end{align}

A sufficient condition that simultaneously satisfies both constraints 
is obtained by requiring $\mu_1$ to lie in the intersection of 
\eqref{eq:app_mu_lmin} and \eqref{eq:app_mu_lmax}. Since the upper 
bound in \eqref{eq:app_mu_lmax} is smaller than that in 
\eqref{eq:app_mu_lmin} (as $\lambda_\text{max} \geq \lambda_\text{min}^+$), 
and the lower bound in \eqref{eq:app_mu_lmin} is smaller than that in 
\eqref{eq:app_mu_lmax} only when 
$\lambda_\text{max}/\lambda_\text{min}^+ \leq \sqrt{\alpha+1} + 1$, 
a simple conservative sufficient condition is obtained by dropping 
the constraint from $\lambda_\text{min}^+$ and retaining only 
\eqref{eq:app_mu_lmax}, which gives \eqref{conver1}. This is 
justified when the condition number 
$\kappa \triangleq \lambda_\text{max}/\lambda_\text{min}^+$ of 
$\boldsymbol{\Phi}^\herm\boldsymbol{\Phi}$ restricted to 
$\mathcal{R}(\boldsymbol{\Phi}^\herm)$ satisfies 
$\kappa \leq \sqrt{\alpha+1} + 1$, which holds in practice when the 
sensing matrix $\boldsymbol{\Phi}$ is well-conditioned on its row space, 
as ensured by a suitable choice of the STAR-RIS control sequence 
$\{\boldsymbol{\Phi}_\text{R}(t)\}_{t=1}^{T_\text{s}}$.
By a local version of the 
Banach fixed-point theorem~\cite{GoebelKirk1990}, the iteration 
$\mathbf{a}^{(\imath+1)} = U_{\mu_1}(\mathbf{a}^{(\imath)})$ converges 
linearly to a fixed point, provided the initial iterate 
$\mathbf{a}^{(0)}$ lies sufficiently close to the ideal solution 
$\mathbf{H}^\text{stack}$.

The local perturbation assumption is satisfied near convergence when the 
noise level $\sigma_e$ is small relative to the singular value gap 
$\Delta\sigma$ of $\mathbf{H}^\text{stack}$, so that the dominant 
rank-$K$ subspace is stable across iterations and 
$\Pi_{\mathcal{R}_K} \approx \widetilde{\Pi}_{\mathcal{R}_K}$. When 
$\sigma_e/\Delta\sigma$ is not negligible, the convergence guarantee is 
approximate, consistent with the mild residual error floor observed in 
Scenario~2 of the numerical results.


\end{document}